\documentclass[twocolumn,prl,superscriptaddress,aps,longbibliography]{revtex4-1}
\usepackage{amsmath}
\usepackage{mathtools}
\usepackage{graphicx}
\usepackage[per-mode=symbol]{siunitx}
\usepackage{ulem}
\DeclareSIUnit\gauss{G}
\DeclareSIUnit\bohr{a_{B}}

\usepackage{hyperref}
\usepackage{bbm}
\usepackage{soul}

\graphicspath{{figures/}}

\usepackage{color}
\definecolor{mygreen}{rgb}{0,0.5,0} 
\definecolor{mygrey}{rgb}{0.5,0.5,0.5} 
\definecolor{myred}{rgb}{0.75,0,0} 
\definecolor{myblue}{rgb}{0,0,0.75} 
\definecolor{mymagenta}{cmyk}{0,1,0,0.12} 
\definecolor{mycyan}{cmyk}{1,0,0,0.12} 
\definecolor{myorange}{rgb}{0.85,0.375,0}  
\definecolor{myviolet}{rgb}{0.5,0.3,1} 
\definecolor{mybrown}{rgb}{0.542969,0.269531, 0.0742188} 

\usepackage{ulem}

\newcommand{\osout}{\bgroup\markoverwith
{\textcolor{myorange}{\rule[0.5ex]{2pt}{0.4pt}}}\ULon}

\begin{document}
\include{./NewCommands}
\newcommand{\microtrap}{microtrap~}
\newcommand{\microtraps}{microtraps~}
\newcommand{\QJPDAcronym}{QJPD}
\newcommand{\supth}{^{(\mathrm{th})}}
\newcommand{\subQJ}{_{\mathrm{QJ}}}
\newcommand{\QE}{\mathrm{QE}}
\newcommand{\subrd}{_\mathrm{rd}}
\newcommand{\subexp}{_\mathrm{exp}}
\newcommand{\subdet}{_\mathrm{det}}
\newcommand{\subcount}{_\mathrm{count}}
\newcommand{\subscat}{_\mathrm{scat}}
\newcommand{\subbg}{_\mathrm{bg}}
\newcommand{\subthr}{_\mathrm{thr}}
\newcommand{\subabs}{_\mathrm{abs}}
\newcommand{\subprobe}{_\mathrm{pr}}
\newcommand{\ndet}{n\subdet}
\newcommand{\Ncount}{N_c}
\newcommand{\ncount}{n_c}
\newcommand{\nfluor}{n_\mathrm{fl}}
\newcommand{\nscat}{n\subscat}
\newcommand{\nbg}{n\subbg}
\newcommand{\nthr}{n\subthr}
\newcommand{\QEJump}{\eta_\mathrm{QJ}}
\newcommand{\QEDet}{\eta\subdet}
\newcommand{\Dtoo}{D\textsubscript{2}}
\newcommand{\D}{\mathrm{D}}
\newcommand{\ND}{\mathrm{ND}}
\newcommand{\QJ}{\mathrm{QJ}}
\newcommand{\NQJ}{\mathrm{NQJ}}

\newcommand{\mytitle}{Quantum jump photodetector for narrowband photon counting with a single atom}
\title{\mytitle}

\newcommand{\ICFO}{ICFO - Institut de Ciencies Fotoniques, The Barcelona Institute of Science and Technology, 08860 Castelldefels, Barcelona, Spain}
\newcommand{\ICREA}{ICREA - Instituci\'{o} Catalana de Recerca i Estudis Avan{\c{c}}ats, 08010 Barcelona, Spain}
\newcommand{\Qruise}{Qruise GmbH, Saarbrücken, D-66113, Germany} 

\author{Laura Zarraoa}
\email{laura.zarraoa@icfo.eu}
\affiliation{\ICFO}
\author{Romain Veyron}
\affiliation{\ICFO}
\author{Tomas Lamich}
\affiliation{\ICFO}
\author{Lorena C. Bianchet}
\affiliation{\ICFO}
\affiliation{\Qruise}
\author{Morgan W. Mitchell}
\email{morgan.mitchell@icfo.eu}
\affiliation{\ICFO}
\affiliation{\ICREA}

\begin{abstract}

Using a single neutral \textsuperscript{87}Rb atom held in an optical trap, and ``quantum jump'' detection of single-photon-initiated state changes, we demonstrate a single-photon quantum jump photodetector (QJPD) with intrinsically narrow bandwidth and strong rejection of out-of-band photons, of interest for detecting weak optical signals in the presence of a strong broadband background. By analyzing fluorescence photon count distributions for the bright and dark states with and without excitation, we measure quantum efficiency  of \SI{2.9+-0.2e-3}{}, a record for single-pass quantum jump production, and {signal-photon-unprovoked ``dark jump'' rate - analogous to the dark count rate of other detectors -} of \SI{3+-10e-3}{jumps\per\second} during passive accumulation plus \SI{4.0+-0.4e-3}{} jumps per readout, orders of magnitude below those of traditional single-photon detectors. Available methods can substantially improve QJPD quantum efficiency, dark {jump rate}, bandwidth, and tunability.

\end{abstract}

\maketitle

\newcommand{\trans}{a}
\newcommand{\emiss}{e}
\newcommand{\prob}{p}
\newcommand{\Hdip}{H_{\rm dip}}
\newcommand{\bmu}{\boldsymbol{\mu}}

Single-photon detection is 
used in many low-light sensing applications, for example laser ranging (LIDAR) \cite{WinkerJAOT2009}, astronomical observation \cite{WalterPASP2020}, and fluorescence microscopy \cite{KonigMAF2020}. Single-photon detection also enables efficient measurement of photon correlations, central to quantum optics \cite{GiustinaPRL2015, ShalmPRL2015} and optical quantum technologies \cite{ZhongS2020}. Most single-photon detectors function by transducing photon-induced processes in solid-state materials, e.g., photoionization from metals \cite{PMT-BeckerHickl}, electron-hole pair production in semiconductors \cite{EMCCD-iKon, CMOS-Retiga, CMOS-ORCA} or heating of superconductors \cite{Dreyling-Eschweiler2015, SNSPD-SingleQuantum}. In solid-state detectors, careful design and fabrication allows high quantum efficiency (QE) \cite{SperlichMST2013, ShaoOQE2021}, i.e., probability of 
detecting the presence of a photon. The use of low-defect materials and cryogenic temperatures can greatly reduce dark counts (DC) \cite{HochbergPRL2019}, i.e., detection events not caused by arriving photons. 

Some single-photon counting applications require, in addition to sensitivity, a strong rejection of background photons. Existing applications include free-space {quantum communications \cite{LoPRL2004, Liao2017,Klop2021,Yang2020}}, and  ``light-shining-through-walls'' searches for dark matter \cite{ChilesPRL2022, SpectorInKimballBook2023}. Emerging applications include near-space and deep-space classical communications \cite{ChenOQE2022} and {spectroscopy of exoplanet atmospheres \cite{RustamkulovN2023, Sliski2023, Crass2020}}. Solid-state detectors, because they are intrinsically broadband, must be complemented by filters if they are to be used in such background-sensitive applications. To be effective against a broad background such as sunlight, such filters must have strong blocking for all frequencies outside a narrow pass-band \cite{ZielinskaOL2012}.

In this context, it is interesting to consider photon counting methods based on the photoresponse of intrinsically-narrowband systems such as atoms. Available atomic systems have optical  bandwidths ranging from a few $\SI{}{\giga\hertz}$ in hot vapor filters \cite{ZielinskaOL2012, ZielinskaOE2014} used for background rejection in LIDAR sky observation \cite{XiaRS2023}\footnote{EIT systems \cite{Zou2017} can have narrower transmission windows within a narrow blocking band. 
\SI{}{\giga\hertz}-bandwidth atomic vapor filters are used in rejection of daylight background \cite{KieferSR2014, XiaRS2023} and can outperform classical filters in some contexts \cite{ YinIEEEPJ}. A relevant measure is the equivalent noise bandwidth \cite{ZielinskaOL2012}.}
to \SI{}{\milli\hertz} in cold atoms on forbidden transitions \cite{BowdenSR2019}. Atomic resonance frequencies are fixed, but a detector based on atomic resonances can in principle detect any frequency through optical frequency conversion, which preserves both photon number and quantum correlations \cite{WangNPJQI2023}. A single atom typically has a low probability of interacting with a single passing photon, but this probability can be raised, in principle to unity, by high numerical aperture (high-NA) focusing \cite{TeyNJP2009, SondermannAQT2020} or optical resonators \cite{NguyenPRA2017}. 

Here we study the use of a single cold, trapped neutral atom with \SI{6}{\mega\hertz} optical bandwidth as a single-photon detector based on ``quantum jump'' (QJ) techniques, i.e., we demonstrate a quantum jump photodetector (\QJPDAcronym). We describe the nature of QE and DC in this system, which resembles CCD \cite{EMCCD-iKon} and CMOS \cite{CMOS-ORCA, CMOS-Retiga} detectors in that it has separated acquisition and readout time windows with distinct DC contributions. We introduce methodology, specific to the \QJPDAcronym{} scenario, for establishing the QE and DC contributions in quantum jump photodetection.
Using a single \textsuperscript{87}Rb atom as a photodetector for \SI{780}{\nano\meter} light, we demonstrate a QE of \SI{2.9+-0.2e-3} (a record for a single atom in free space), a {dark jump rate} of \SI{3+-10e-3}{jumps\per\second}, consistent with zero and limited by measurement statistics, and a readout contribution of {\SI{3.1+-0.04e-2}{} erroneous counts} per readout.  {These observed dark count contributions}
are already competitive with any non-cryogenic detector, and could benefit from several proven atomic and optical techniques. As noted already, the QE can in principle be raised close to unity through various enhancement techniques. 

\begin{figure*}[t]
    \centering
\includegraphics[width=0.9\textwidth]{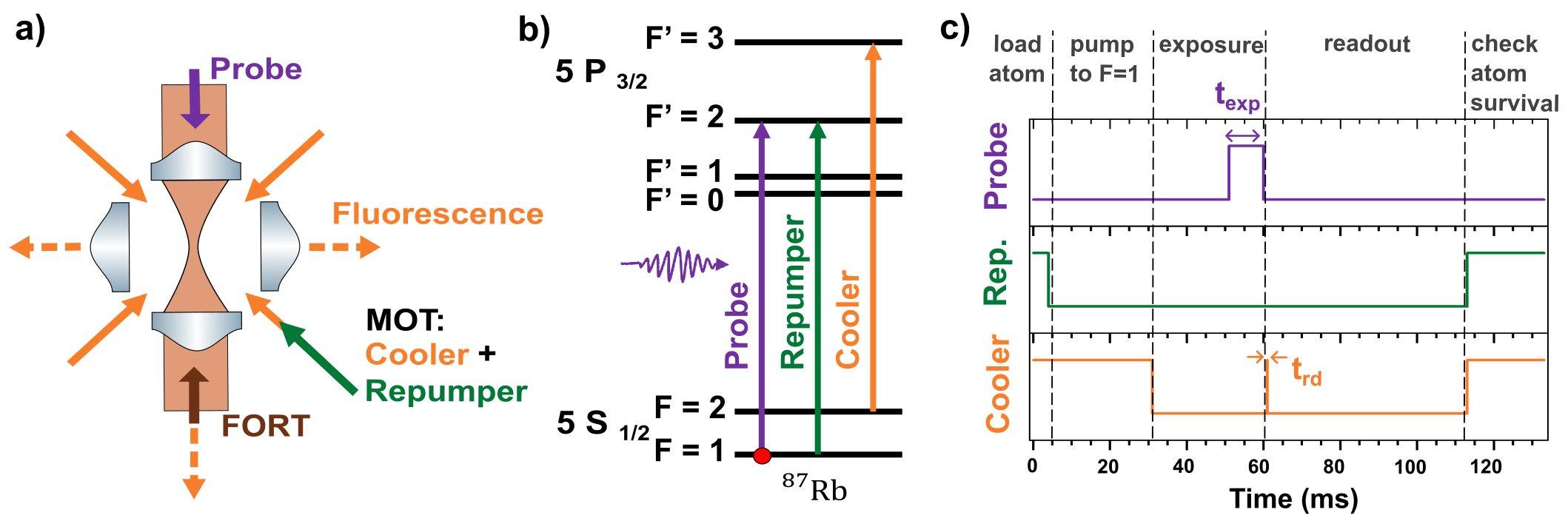}
    \caption{Overview of the \QJPDAcronym{} operation. (a) Experimental set-up consisting of four high numerical aperture lenses in a Maltese cross configuration. (b) Lambda system in which the $^{87}$Rb \QJPDAcronym{}  operates. (c) Experimental sequence used for quantum jump detection: an atom is loaded into the trap, prepared in the $F=1$ ground state, exposed to probe photons during an exposure time $t\subexp$, and has its state ``read out'' by exposure to cooler light for time $t\subrd$. Exposure and readout steps (grey band) can be repeated for continuous, low dead-time acquisition, limited only by the survival of the atom in the trap. For the characterization experiments presented here, these steps were not repeated.  MOT light including repumper is then applied to check if the atom is still in the trap. 
 }
    \label{fig:SetupLevelsTiming}
\end{figure*}

QJs, first studied experimentally in single trapped ions \cite{NagourneyPRL1986}, are abrupt transitions of an atom between states or groups of states with observably different fluorescence behavior. We illustrate this with the \textsuperscript{87}Rb \Dtoo{} transitions, from $5\mathrm{S}_{1/2}~ F$ (ground states) to $5\mathrm{P}_{3/2}~ F'$  (excited states), as shown in Fig.~\ref{fig:SetupLevelsTiming}(b). When illuminated with laser light tuned to the closed $F=2 \rightarrow F'=3$  ``cooler'' transition, the $F=2$ ground state continuously emits resonance fluorescence.
Under the same illumination, the $F=1$ ground state produces negligible fluorescence, because of the larger detuning from any allowed transition. Absorption of a single ``probe'' photon tuned to the $F=1 \rightarrow F'=2$ transition, followed by spontaneous emission to the $F=2$ ground state, thus induces a transition between a low-  and high-fluorescence condition, which can be detected by collecting and counting the resonance fluorescence photons. In effect, the probe photon, by causing a change of atomic state, triggers a cascade of resonance fluorescence events.  A similar system has made precise, sub-wavelength measurement of off-resonance light intensity  \cite{Bianchet2022}, at a flux of $\sim \SI{e15}{photons\per\second}$ and with optical bandwidth $\sim \SI{e14}{\hertz}$. Here we study the use of this system to count resonant photons, with flux $\sim \SI{1e0}{photon\per\second}$, bandwidth $\SI{6e6}{\hertz}$, and statistics suited to detection of individual events.

A practical sequence is shown in Fig.~\ref{fig:SetupLevelsTiming}(c): the atom is initialized in state $F=1$, exposed to probe light during an exposure time window and then read out by illumination with cooler light.
During exposure, an incoming probe photon promotes the atom to $F'=2$ with probability $\eta\subabs$, determined by the matching of the incoming photon properties (spatial, spectral, and polarization) to the allowed $F=1\rightarrow F'=2$ transitions \cite{TeyNJP2009, AljunidPRL2013}. 
From the $F'=2$ state, the atom falls to the $F=1$ {state with probability $q$, or to $F=2$ with probability $1-q$, determined by the branching ratio. In our case, $q=1/2$.}

If the atom falls to the initial state, in our example $F=1$, the absorption event has failed to produce a quantum jump. The efficiency to {reach the excited state is $q\eta\subabs$, in our example $\eta\subabs/2$. Meanwhile, for a natural-lifetime limited transition and a single-pass interaction of the probe photon with the atom, the efficiency for producing a jump is $\eta\subQJ$, which is upper-bounded by $q$, so that $\eta\subQJ \le q(1-q)$, which saturates at $1/4$ for a 1:1 branching ratio, as is the case for our system.} 

During the readout, an atom in $F=2$ 
stochastically scatters photons on the $F = 2 \leftrightarrow F'=3$ closed transition before eventually scattering a photon via $F'=2$ or $F'=1$ to fall to $F=1$, at which point the fluorescence ceases. {The scattered photons are collected and detected with net efficiency $\eta\subdet$ by single-photon avalanche diodes (SPADs). The observed counts are $\ncount = \nfluor+\nbg$, where $\nbg$ is the detector background counts and $\nfluor$ is the number of detected fluorescence photons. When the atom starts in F=1 (or F=2) $\ncount$ has distribution
\begin{eqnarray}
\label{eq:PncF=1}
P(\ncount|F=1) &=& \mathcal{P}_{\mu\subbg}(\ncount)
\\
\label{eq:PncF=2}
P(\ncount|F=2) &=& [\mathcal{P}_{\mu\subbg}(\nbg)  * P_{\eta\subdet}(\nfluor|F=2)](\ncount), \hspace{6mm}
\end{eqnarray}
where we assume a Poissonian background $\mathcal{P}_{\mu}(n) \equiv \exp[-\mu] \mu^{n}/n!$ with mean $\mu$. For F=2, $\ncount$ distribution is the convolution ($*$) of the background and $P_{\eta\subdet}(\nfluor|F=2)$, the probability of detecting $\nfluor$ fluorescence photons} for a {given efficiency $\eta\subdet$. A specific model for $P_{\eta\subdet}(\nfluor|F=2)$, appropriate for homogenously broadened systems, is given in the  Appendix.
}

We implement a \QJPDAcronym{} using a single neutral \textsuperscript{87}Rb atom in a strongly-focused far-off-resonance trap (FORT) surrounded by high-NA lenses. The experimental system is shown schematically in Fig.~\ref{fig:SetupLevelsTiming}(a), and is described in detail in \cite{BrunoOE2019, BianchetORE2021}.  
In the following, all laser detunings are given with respect to the natural \textsuperscript{87}Rb transitions {in units of the natural decay rate $\Gamma_0$} . The atom experiences dynamical Stark shifts induced by the FORT \cite{SimonCoopLightshift2017, Bianchet2022}: at 
trap center, the light shift of the $F=1 \rightarrow F'=2$ transition is $3.3\Gamma_0$ and scalar, with negligible tensor contribution. For the cooler transition, the tensorial polarisability leads to Zeeman-state-dependent excited state light shifts, ranging from $2.6 \Gamma_0$ ($m_{F'}=\pm3$) to $4.2 \Gamma_0$ ($m_{F'}=0$). Lenses surround the atom in the horizontal plane. The FORT and probe polarizations, as well as an applied bias magnetic field of \SI{180}{\milli\gauss}, are vertical, which we take as the quantization axis.

\newcommand{\FigHeight}{5.75cm}
\newcommand{\FigHeightShort}{5.25 cm}

\begin{figure}[t!]
    \centering
    \includegraphics[height=\FigHeightShort]{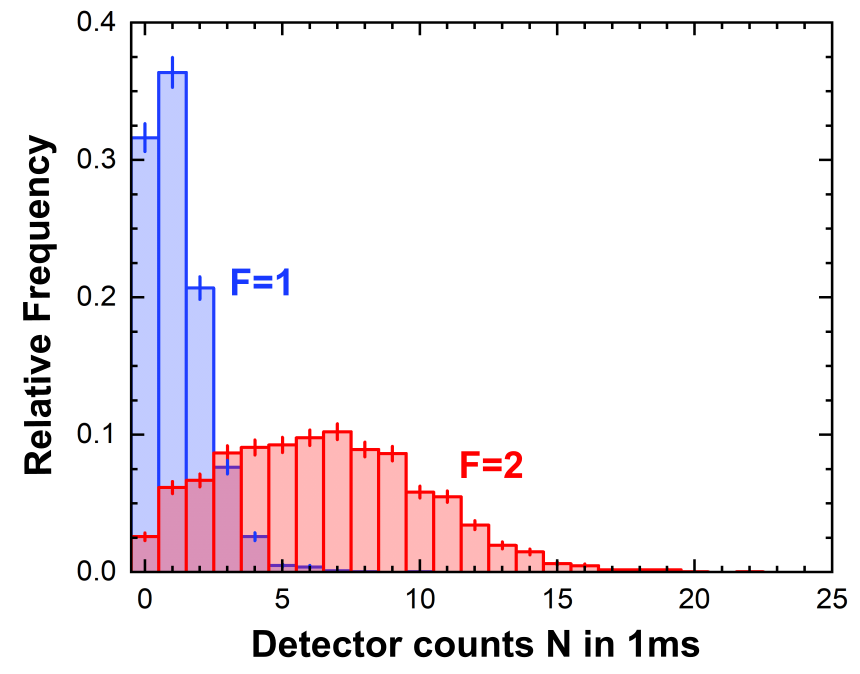}
    \caption{Experimental characterization of the state-dependent fluorescence of the atom: histograms show the number of counts $\ncount$ collected when an atom, initially prepared in the $F=1$ (blue) or $F=2$ (red) ground state, is illuminated for \SI{1}{\milli\second} by cooler light. $\ncount$ is obtained by summing SPAD detection events from light collected by two high-NA lenses, one along the trap axis and one transverse to the axis.
      }
    \label{fig:DetectorCountHistograms}
\end{figure}

The atom is loaded from a three-dimensional magneto-optical trap (MOT) into the FORT, with  depth \SI{787+-2}{\micro\kelvin}. The atom is cooled by polarization gradient cooling for \SI{4}{\milli\second} at a detuning of $3.75 \Gamma_0$ to the red of the  $F=2 \rightarrow F'=3$ transition and intensity of \SI{42}{\watt\per\meter\squared} per MOT beam. We first measure the conditional count distributions of Eqs.~(\ref{eq:PncF=1}) and (\ref{eq:PncF=2}), by   optically pumping the atom to $F=1$ or $F=2$, respectively, and reading out the fluorescence. As illustrated in Fig.~\ref{fig:SetupLevelsTiming}(c), the atom is prepared in the $F=1$ ground state by a \SI{27}{\milli\second}  duration pulse tuned $3\Gamma_0$ to the red of the  $F=2 \rightarrow F'=3$ transition. The $F=2$ preparation (not shown in  Fig.~\ref{fig:SetupLevelsTiming}(c)) is performed by illuminating with a \SI{10}{\milli\second}  pulse tuned $5\Gamma_0$ to the blue of the  $F=2 \rightarrow F'=3$ transition. 

The trap depth is then lowered to \SI{558+-2}{\micro\kelvin}  to reduce the FORT-induced differential light shifts on the $F'=3$ state. At this stage the atom temperature is \SI{15+-2}{\micro\kelvin}. The atom is then illuminated by a readout pulse of {cooler light along the MOT beams} with duration $t\subrd = \SI{1}{\milli\second}$, intensity \SI{20}{\watt/\meter\squared} per beam, and tuned $0.5\Gamma_0$ to the blue of the $F=2 \rightarrow F'=3$ transition. Fluorescence photons during this time are collected through $\mathrm{NA} = 0.5$ aspheric lenses and counted by SPADs (Excelitas SPCM-AQ4C). Their histograms, relative frequencies  $f(\ncount|F=1)$ and $f(\ncount|F=2)$, are shown in Fig.~\ref{fig:DetectorCountHistograms}, and are well approximated by a Poisson distribution with $\mu = \SI{1.146+-0.002}{counts}$, and by a Gaussian with mean \SI{5.9+-0.1}{counts} and variance \SI{17.6+-0.8}{counts} squared,  respectively.  

From Eqs.~(\ref{eq:PncF=1}) and (\ref{eq:PncF=2}), $P(\ncount|F)$ is nonvanishing for any $\ncount$ and $F$, meaning we cannot unambiguously infer $F$ from $\ncount$ and must resort to decision theory \cite{PetersonBook2017}. Because the QJ increases the fluorescence rate of the atom, it is natural to define detection (D) of the QJ as $\ncount > \nthr$ and non-detection (ND) as $\ncount  \le \nthr$, where $\nthr$ is a threshold to be chosen in function of the cost of false negatives $\epsilon_\mathrm{FN} \equiv P(\ncount \le \nthr|F=2)$, which decrease QE, and false positives $\epsilon_\mathrm{FP} \equiv P(\ncount > \nthr|F=1)$, which contribute to readout noise. 
Although other choices may better suit specific applications, here we choose $\nthr = 4$, which maximizes readout fidelity $\mathcal{F} \equiv 1 - (\epsilon_\mathrm{FP} + \epsilon_\mathrm{FN})/2$ as defined by Fuhrmanek et al. \cite{Fuhrmanek2011}, to give {$\mathcal{F}=0.93$.}

Given $n\subthr$ and  $P(\ncount|F)$, and thus $P(\D|F)$, the probability of a quantum jump $P(\mathrm{QJ})$ can be estimated 
as:
\begin{eqnarray}
P(\QJ) &=& \frac{P(\D) - \epsilon_\mathrm{FP}}{1-\epsilon_\mathrm{FP} - \epsilon_\mathrm{FN}}.
\label{eq:PQJFromPD}
\end{eqnarray}
The uncertainty in this inference can be calculated by error propagation: For the binomial outcome D/ND, the mean-squared error of $P(\QJ)$ is:
\begin{eqnarray}
\mathrm{MSE}[P(\QJ)] &=& \frac{1}{(1-\epsilon_\mathrm{FP} - \epsilon_\mathrm{FN})^2} \frac{P(\mathrm{D})[1-P(\mathrm{D})]}{N_\mathrm{runs}}\label{eq:ErrorPQJ}, \hspace{3mm}
\end{eqnarray}
where $N_\mathrm{runs}$ is the number of repetitions of the experiment. 
Eqs.~(\ref{eq:PQJFromPD}) and~(\ref{eq:ErrorPQJ}) will be used to estimate the number of QJs and its uncertainty when characterizing quantum efficiency and dark counts, e.g., for Fig.~\ref{fig:QuantumJumpSaturation}. See also Hemmerling \textit{et al.} \cite{Hemmerling2012}.
  
\newcommand{\nprobebar}{\bar{n}\subprobe}
\begin{figure}[t]
\centering
\includegraphics[height=\FigHeight]{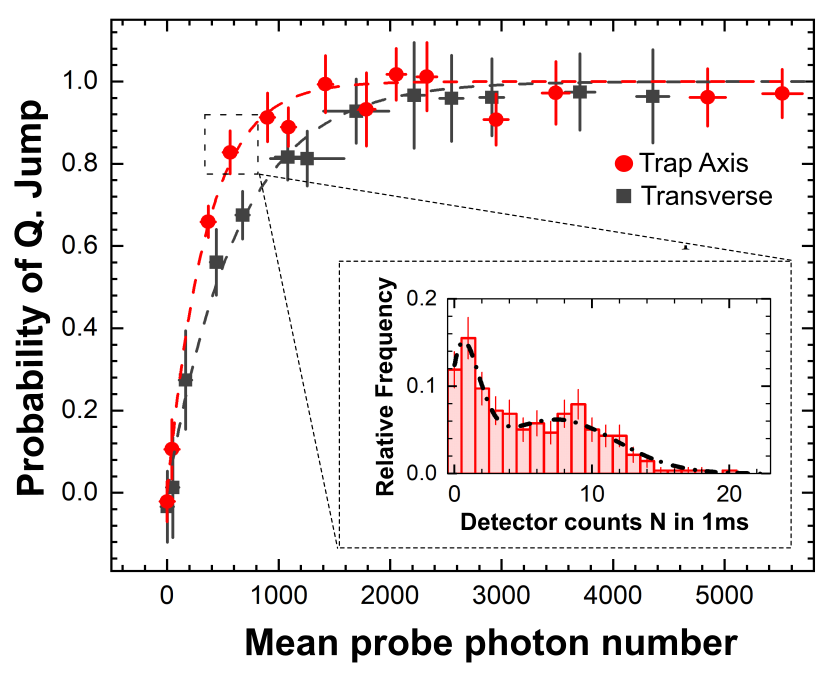}
\caption{Probabilities $P(\QJ|\nprobebar)$ of a quantum jump versus mean number of probe photons $\nprobebar$ illuminating the atom. Symbols show $P(\QJ|\nprobebar)$, calculated using Eq.~(\ref{eq:PQJFromPD}). {$\QEJump$ is calculated from light collected from one lens along the trap axis and one lens along the transverse axis for the case of probe illumination along the trap axis, and from two opposite lenses along the trap axis for the case of probe illumination in the transverse axis.}
The \textit{y} axis error bars show $\pm 1 \sigma$ uncertainty computed using Eq.~(\ref{eq:ErrorPQJ}). The \textit{x} axis error bars show combined statistical (from fluctuation during acquisition) and systematic (from probe power calibration) uncertainty in $\nprobebar$. Dashed lines show fits with Eq.~(\ref{eq:SaturationCurve}) with results: $\QEJump = \SI{2.9+-0.2e-3}{}$ for illumination along the trap axis (red), and $\QEJump = \SI{1.6+-0.06e-3}{}$ for illumination {transversal} to the trap axis (black).  Inset shows a histogram  of $\ncount$ (red bars) for probe illumination with $\nprobebar = \SI{570+-40}{photons}$ and the fit (black dashed line) with Eq.~(\ref{eq:Fit_B}) to obtain $P(\QJ)$. }
\label{fig:QuantumJumpSaturation}
\end{figure}

Because the QJPD detects photons in two steps, namely the $F=1 \rightarrow F=2$ state change (the QJ itself) and the subsequent observation by resonance fluorescence, we can identify two QEs: the QJ efficiency $\eta\subQJ$ and the detection efficiency $\eta\subdet$. To measure these, we use the sequence shown in {Fig.~\ref{fig:SetupLevelsTiming}(c)} with pulse durations and power levels as given above, with the addition of a strongly-attenuated laser probe pulse of duration $t\subexp = \SI{10}{\milli\second}$, {blue detuned $3.3 \Gamma_0$} from the $F=1 \rightarrow F'=2$ transition, focused onto the atom by one of the high-NA lenses and of variable mean photon number $\nprobebar$. To monitor the probe photon number, the probe light is split by a 50:50 fiber beamsplitter, with one output going directly to one SPAD, and the other to the QJPD. The SPAD is calibrated against power measurements before and after the vacuum chamber, assuming equal window losses.

For each probe power setting, 300 runs of the prepare-probe-readout sequence are made to build a histogram of $\ncount$, as illustrated in the inset of Fig. \ref{fig:QuantumJumpSaturation}.  This is fitted with
\begin{eqnarray}
\label{eq:Fit_B}
P(\ncount|\nprobebar)&=&P(\ncount|F=1)P(F=1|\nprobebar) \nonumber \\ & & + P(\ncount|F=2)P(F=2|\nprobebar), 
\end{eqnarray}
a weighted sum of the conditional probabilities of Eqs.~(\ref{eq:PncF=1}) and (\ref{eq:PncF=2}) with a single fit parameter $P(\mathrm{QJ}|\nprobebar) \equiv P(F=2|\nprobebar) = 1-P(F=1|\nprobebar)$. The resulting values of $\eta\subQJ(\nprobebar)$, both for illumination along the trap axis and transverse to the trap axis, are shown in Fig. \ref{fig:QuantumJumpSaturation}, and are well described by single-parameter saturation curves
\begin{eqnarray}
P(\mathrm{QJ}|\nprobebar) &=&  1 - \exp[{-\QEJump \nprobebar}],
\label{eq:SaturationCurve}
\end{eqnarray}
with quantum jump efficiency $\QEJump = \SI{2.9+-0.2e-3}{}$ for illumination along the trap axis, and $\QEJump = \SI{1.6+-0.06e-3}{}$ for illumination orthogonal to the trap axis. 
The higher on-axis coupling efficiency reflects the elongated shape of the atomic probability distribution. 
These $\QEJump$ values are the highest reported quantum jump probabilities for free-space single-photon level coupling to either neutral atoms or to ions \cite{VolzWeberSchlenkEtAl2006}. 

\newcommand{\subDJ}{_\mathrm{DJ}}
\newcommand{\subRJ}{_\mathrm{RJ}}
\newcommand{\subFP}{_\mathrm{FP}}
\newcommand{\subRE}{_\mathrm{RE}}
\newcommand{\subDC}{_\mathrm{DC}}

{To understand DCs in the \QJPDAcronym{}, we study the two distinct sources of detections in the absence of incoming probe photons. First, there can be a quantum jump during the exposure phase, e.g., due to scattering of FORT light.
Second, there can be false positives in the readout process. These two sources play the same roles, respectively, as ``dark current'' and ``readout noise'' in a CCD \cite{EMCCD-iKon} or CMOS sensor \cite{CMOS-ORCA}. By analogy, we call these ``dark jumps'' (DJs) and  ``readout errors'' (REs). 
The RE itself has two contributions: a ``read jump'' (RJ) and a ``fluorescence false positive'' (FP). A RJ is a QJ during the readout pulse, e.g., due to scattering of cooler light. A FP is an above-threshold SPAD count in the absence of a QJ, due, e.g., to background light or SPAD thermal firings. We write the corresponding probabilities $\epsilon_{\text{RE}}$, $ \epsilon\subRJ$ and $\epsilon\subFP$.}
The probability of a read jump grows at rate $r\subRJ$ during the read time $t_{\text{rd}}$ to give $\epsilon\subRJ = 1 - \text{exp}[-t\subrd r\subRJ] \approx t\subrd r\subRJ$
{The dark counts per acquisition are thus $\epsilon\subDC = t\subexp r\subDJ + t\subrd r\subRJ + \epsilon_\mathrm{FP}$.} 

For our chosen threshold of $\nthr = 4$, $\epsilon_\mathrm{FP}=0.027$. We can experimentally quantify {$r\subDJ$ and $\epsilon\subRJ$. Both these quantities imply a QJ rate that depends on the time between readout, and thus on the time resolution of the acquisition. They are independent of the fluorescence detection hardware, and thus represent a more fundamental performance limit set by the atomic level structure.}

{
To quantify the {dark jumps,} the atom is prepared in $F=1$ and remains trapped, illuminated only by the trapping light for time $t\subexp$, after which it is read out with a \SI{1}{\milli\second} pulse to find $\ncount$ and infer D/ND. We make 500 runs each for $t\subexp=\SI{0.01}{\second}, \SI{1}{\second}, \SI{2}{\second}$ and \SI{3}{\second}, limited by the trap lifetime due to background gas collisions. Runs in which the atom is not present after the sequence are discarded. We compute $P$(QJ) using Eq.~(\ref{eq:PQJFromPD}) with $\epsilon_{\text{FP}}$ and $\epsilon_{\text{FN}}$ computed from measured histograms with no atom and with an atom in $F=2$. Linear regression of the $P$(QJ) versus $t\subexp$ finds a dark jump rate of \SI{3+-10e-3}{jumps\per\second}. This result, consistent with zero, is limited by the false positive rate
and the number of atoms tested. It places an upper limit on the dark jump contribution to the QJPD DC.

The {RJ} rate $r\subRJ$ is quantified in a similar way: We prepare the atom in $F=1$, wait \SI{10}{\milli\second} (equal to $t\subexp$ when measuring QE), apply a readout pulse of duration $t\subrd$, record the resulting $\ncount$ and infer D/ND by comparison to $\nthr$ as defined above. For each $t\subrd$ value, 500 runs are acquired and runs with lost atoms are discarded. We compute $P$(QJ) as in the previous paragraph. Linear regression of this $P$(QJ) versus $t\subrd$ finds the RJ rate of \SI{4+-0.4}{jumps\per\second} of readout duration. Our readout pulse is \SI{1}{\milli\second} in duration, {giving $\epsilon\subRE = t\subrd r\subRJ + \epsilon_\mathrm{FP} = \SI{4+-0.4e-3}{} + 0.027 = \SI{3.1+-0.04e-2}{}$ errors per read. }

}

With a \SI{1}{\second} exposure window, and thus  with a nearly \SI{1}{\hertz} readout rate, the net DC rate is far below those of commercially-available single-photon detectors, including photomultipliers \cite{PMT-BeckerHickl}, semiconductor devices \cite{CMOS-ORCA, CMOS-Retiga, EMCCD-iKon, CMOS-Retiga,CMOS-ORCA, SAPDs}, and superconducting devices \cite{SNSPD-SingleQuantum, SiNWs_Psyche} but exceeds those of some research superconducting detectors \cite{Dreyling-Eschweiler2015, HochbergPRL2019, ChilesPRL2022}.  Several proven techniques could improve \QJPDAcronym{} performance. Readout noise can be reduced by Zeeman-state pumping during readout \cite{Fuhrmanek2011}, using SNSPD in place of SPADs, and improved vacuum to extend trap lifetime. These would increase the statistical distance between $P(\ncount|F=1)$ and $P(\ncount|F=2)$ and thus reduce false positives. Dark {jumps}, too small to measure in this experiment, can be produced by off-resonance optical excitation by FORT light or thermal photons, or by magnetic excitation of the ground-state hyperfine transition. We estimate the FORT to be the strongest contributor, with thermal and magnetic background being negligible. A blue-detuned FORT or lattice, which traps the atom at an intensity minimum, could thus reduce the dark {jumps}. A magic wavelength optical trap could in addition reduce optical linewidth broadening \cite{ZhangEFTFIFCS2022}. Non-optical trapping, e.g., for ions \cite{PiroNP2011}, similarly escapes these effects.
QE can be increased by initializing the atom in a specific $m_F$ state \cite{PiroNP2011, SpechtN2011}, and by optical methods including strong beamshaping \cite{SondermannAQT2020} and cavity enhancement \cite{KuhnPRL2002}. Cavity-QED methods can moreover escape the $\eta\subQJ \le 1/4$ efficiency limit in driving Raman transitions such as $F=1\rightarrow F=2$ \cite{SpechtN2011}.

In conclusion, we have demonstrated a quantum jump photodetector (QJPD) with an intrinsically narrow \SI{6}{\mega\hertz} response bandwidth. We introduce and demonstrate methodology for measuring the quantum efficiency and the two dark count contributions of a QJPD, and find dark counts below those of any commercially-available single-photon detector. The QJPD technique may be interesting for the growing number of applications in communications and fundamental science that require high sensitivity and strong background rejection through frequency discrimination. Several proven atomic and optical technologies could be applied to reach different wavelength ranges, narrower bandwidths, higher quantum efficiency, and lower dark counts.  

\begin{acknowledgements}
Acknowledgements: We thank V. Prakash and N. Alves for helpful discussions on the development and implementation of the experiments. We further thank V. Prakash the careful reading of the manuscript. Funded by the European Union (ERC, Field-SEER, 101097313 and QUANTIFY, 101135931), NextGenerationEU/PRTR, and by the Spanish Ministry of Science MCIN: project SAPONARIA (PID2021-123813NB-I00) and ``Severo Ochoa'' Center of Excellence CEX2019-000910-S, Departament de Recerca i Universitats de la Generalitat de Catalunya grant No. 2021 SGR 01453, Fundaci\'{o} Privada Cellex, Fundaci\'{o} Mir-Puig. LZ acknowledges the “Presidencia de la Agencia Estatal de Investigación” grant Ref. PRE2020-094392. TL acknowledges Marie Sk\l{}odowska-Curie grant agreement No 847517.  Views and opinions expressed are those of the authors only and do not necessarily reflect those of the European Union or the European Research Council Executive Agency. Neither the European Union nor the granting authority can be held responsible for them.
\end{acknowledgements}

\noindent
\textit{Author contributions} - 
L.C.B., L.Z., and M.W.M. developed the QJPD concept. L.C.B. performed preliminary experiments and theoretical analysis not reported here. L.Z., R.V. and T.L. performed the experiments reported and analysed the data. R.V. and M.W.M. developed the analysis approach and supervised the study.


\noindent
\textit{Appendix} -- Minimal stochastic model for $P(\ncount|F=2)$: The atom starts in $F=2$, with probability $p$ to go to $F=2$, and probability $1-p$ to go to $F=1$, when scattering a photon. The probability to scatter exactly $s$ photons before arriving to the $F=1$ state and going dark is  exponential: $P(\nscat=s) =  p^{s-1}(1-p)$ for $s \ge 1$ and zero otherwise. If the efficiency from scattering to detection is $\eta$, the probability of detecting $\ndet=d$ photons out of $s$ scattered photons is binomial: $P(\ndet=d | \nscat=s)= \eta^{d} (1-\eta)^{(s-d)} \tbinom{s}{d}$. Summing these conditional distributions, weighted by the probability to scatter $s$ photons, we have the unconditional probability of detecting $d$ photons:  $P(\ndet=d) = \sum_{s=d}^\infty P(\ndet=d | \nscat=s) P(\nscat=s) = p^{d-1}(1-p) \eta^d/ (1 - p + \eta p)^{d+1}$. We convolve this with the Poisson distribution for the background counts to find the distribution for $\ncount \equiv \ndet + \nbg$: 
\begin{eqnarray}
P(\ncount|F=2) &=& 
\sum_{d=0}^{\ncount} P(\ndet=d) \frac{e^{-\mu} \mu^{\ncount - d}}{(\ncount - d)!} \\ & = & 
 \frac{\mu^{\ncount}}{\ncount!p}x e^{x} E_{-\ncount}[ x + \mu],
 \label{eq:SimpleMarkovModelResult}
\end{eqnarray}
where $x \equiv \mu(1-p)/\eta p$ and $E_n[z]$ is the exponential integral function. $P(\ncount|F=2)$ can also be expressed in terms of the incomplete Gamma function. The fact that Eq.~(\ref{eq:SimpleMarkovModelResult}) does not fit well the $f(\ncount|F=2)$ of Fig.~\ref{fig:DetectorCountHistograms} suggests that inhomogeneity, e.g., in the response of different Zeeman states  or to different cooler polarizations, plays an important role in determining $P(\ncount|F=2)$. For this reason, we use a heuristic (Gaussian) distribution in fitting $f(\ncount|F=2)$.

%


\begin{thebibliography}{53}%
\makeatletter
\providecommand \@ifxundefined [1]{%
 \@ifx{#1\undefined}
}%
\providecommand \@ifnum [1]{%
 \ifnum #1\expandafter \@firstoftwo
 \else \expandafter \@secondoftwo
 \fi
}%
\providecommand \@ifx [1]{%
 \ifx #1\expandafter \@firstoftwo
 \else \expandafter \@secondoftwo
 \fi
}%
\providecommand \natexlab [1]{#1}%
\providecommand \enquote  [1]{``#1''}%
\providecommand \bibnamefont  [1]{#1}%
\providecommand \bibfnamefont [1]{#1}%
\providecommand \citenamefont [1]{#1}%
\providecommand \href@noop [0]{\@secondoftwo}%
\providecommand \href [0]{\begingroup \@sanitize@url \@href}%
\providecommand \@href[1]{\@@startlink{#1}\@@href}%
\providecommand \@@href[1]{\endgroup#1\@@endlink}%
\providecommand \@sanitize@url [0]{\catcode `\\12\catcode `\$12\catcode `\&12\catcode `\#12\catcode `\^12\catcode `\_12\catcode `\%12\relax}%
\providecommand \@@startlink[1]{}%
\providecommand \@@endlink[0]{}%
\providecommand \url  [0]{\begingroup\@sanitize@url \@url }%
\providecommand \@url [1]{\endgroup\@href {#1}{\urlprefix }}%
\providecommand \urlprefix  [0]{URL }%
\providecommand \Eprint [0]{\href }%
\providecommand \doibase [0]{http://dx.doi.org/}%
\providecommand \selectlanguage [0]{\@gobble}%
\providecommand \bibinfo  [0]{\@secondoftwo}%
\providecommand \bibfield  [0]{\@secondoftwo}%
\providecommand \translation [1]{[#1]}%
\providecommand \BibitemOpen [0]{}%
\providecommand \bibitemStop [0]{}%
\providecommand \bibitemNoStop [0]{.\EOS\space}%
\providecommand \EOS [0]{\spacefactor3000\relax}%
\providecommand \BibitemShut  [1]{\csname bibitem#1\endcsname}%
\let\auto@bib@innerbib\@empty
\bibitem [{\citenamefont {Winker}\ \emph {et~al.}(2009)\citenamefont {Winker}, \citenamefont {Vaughan}, \citenamefont {Omar}, \citenamefont {Hu}, \citenamefont {Powell}, \citenamefont {Liu}, \citenamefont {Hunt},\ and\ \citenamefont {Young}}]{WinkerJAOT2009}%
  \BibitemOpen
  \bibfield  {author} {\bibinfo {author} {\bibfnamefont {David~M.}\ \bibnamefont {Winker}}, \bibinfo {author} {\bibfnamefont {Mark~A.}\ \bibnamefont {Vaughan}}, \bibinfo {author} {\bibfnamefont {Ali}\ \bibnamefont {Omar}}, \bibinfo {author} {\bibfnamefont {Yongxiang}\ \bibnamefont {Hu}}, \bibinfo {author} {\bibfnamefont {Kathleen~A.}\ \bibnamefont {Powell}}, \bibinfo {author} {\bibfnamefont {Zhaoyan}\ \bibnamefont {Liu}}, \bibinfo {author} {\bibfnamefont {William~H.}\ \bibnamefont {Hunt}}, \ and\ \bibinfo {author} {\bibfnamefont {Stuart~A.}\ \bibnamefont {Young}},\ }\bibfield  {title} {\enquote {\bibinfo {title} {Overview of the {CALIPSO} mission and {CALIOP} data processing algorithms},}\ }\href {\doibase 10.1175/2009JTECHA1281.1} {\bibfield  {journal} {\bibinfo  {journal} {Journal of Atmospheric and Oceanic Technology}\ }\textbf {\bibinfo {volume} {26}},\ \bibinfo {pages} {2310 -- 2323} (\bibinfo {year} {2009})}\BibitemShut {NoStop}%
\bibitem [{\citenamefont {Walter}\ \emph {et~al.}(2020)\citenamefont {Walter}, \citenamefont {Fruitwala}, \citenamefont {Steiger}, \citenamefont {Bailey}, \citenamefont {Zobrist}, \citenamefont {Swimmer}, \citenamefont {Lipartito}, \citenamefont {Smith}, \citenamefont {Meeker}, \citenamefont {Bockstiegel}, \citenamefont {Coiffard}, \citenamefont {Dodkins}, \citenamefont {Szypryt}, \citenamefont {Davis}, \citenamefont {Daal}, \citenamefont {Bumble}, \citenamefont {Collura}, \citenamefont {Guyon}, \citenamefont {Lozi}, \citenamefont {Vievard}, \citenamefont {Jovanovic}, \citenamefont {Martinache}, \citenamefont {Currie},\ and\ \citenamefont {Mazin}}]{WalterPASP2020}%
  \BibitemOpen
  \bibfield  {author} {\bibinfo {author} {\bibfnamefont {Alexander~B.}\ \bibnamefont {Walter}}, \bibinfo {author} {\bibfnamefont {Neelay}\ \bibnamefont {Fruitwala}}, \bibinfo {author} {\bibfnamefont {Sarah}\ \bibnamefont {Steiger}}, \bibinfo {author} {\bibfnamefont {John~I.}\ \bibnamefont {Bailey}}, \bibinfo {author} {\bibfnamefont {Nicholas}\ \bibnamefont {Zobrist}}, \bibinfo {author} {\bibfnamefont {Noah}\ \bibnamefont {Swimmer}}, \bibinfo {author} {\bibfnamefont {Isabel}\ \bibnamefont {Lipartito}}, \bibinfo {author} {\bibfnamefont {Jennifer~Pearl}\ \bibnamefont {Smith}}, \bibinfo {author} {\bibfnamefont {Seth~R.}\ \bibnamefont {Meeker}}, \bibinfo {author} {\bibfnamefont {Clint}\ \bibnamefont {Bockstiegel}}, \bibinfo {author} {\bibfnamefont {Gregoire}\ \bibnamefont {Coiffard}}, \bibinfo {author} {\bibfnamefont {Rupert}\ \bibnamefont {Dodkins}}, \bibinfo {author} {\bibfnamefont {Paul}\ \bibnamefont {Szypryt}}, \bibinfo {author} {\bibfnamefont {Kristina~K.}\ \bibnamefont {Davis}}, \bibinfo {author}
  {\bibfnamefont {Miguel}\ \bibnamefont {Daal}}, \bibinfo {author} {\bibfnamefont {Bruce}\ \bibnamefont {Bumble}}, \bibinfo {author} {\bibfnamefont {Giulia}\ \bibnamefont {Collura}}, \bibinfo {author} {\bibfnamefont {Olivier}\ \bibnamefont {Guyon}}, \bibinfo {author} {\bibfnamefont {Julien}\ \bibnamefont {Lozi}}, \bibinfo {author} {\bibfnamefont {Sebastien}\ \bibnamefont {Vievard}}, \bibinfo {author} {\bibfnamefont {Nemanja}\ \bibnamefont {Jovanovic}}, \bibinfo {author} {\bibfnamefont {Frantz}\ \bibnamefont {Martinache}}, \bibinfo {author} {\bibfnamefont {Thayne}\ \bibnamefont {Currie}}, \ and\ \bibinfo {author} {\bibfnamefont {Benjamin~A.}\ \bibnamefont {Mazin}},\ }\bibfield  {title} {\enquote {\bibinfo {title} {The {MKID} exoplanet camera for {Subaru} {SCExAO}},}\ }\href {\doibase 10.1088/1538-3873/abc60f} {\bibfield  {journal} {\bibinfo  {journal} {Publications of the Astronomical Society of the Pacific}\ }\textbf {\bibinfo {volume} {132}},\ \bibinfo {pages} {125005} (\bibinfo {year} {2020})}\BibitemShut
  {NoStop}%
\bibitem [{\citenamefont {K\"{o}nig}(2020)}]{KonigMAF2020}%
  \BibitemOpen
  \bibfield  {author} {\bibinfo {author} {\bibfnamefont {Karsten}\ \bibnamefont {K\"{o}nig}},\ }\bibfield  {title} {\enquote {\bibinfo {title} {Review: Clinical in vivo multiphoton {FLIM} tomography},}\ }\href {\doibase 10.1088/2050-6120/ab8808} {\bibfield  {journal} {\bibinfo  {journal} {Methods and Applications in Fluorescence}\ }\textbf {\bibinfo {volume} {8}},\ \bibinfo {pages} {034002} (\bibinfo {year} {2020})}\BibitemShut {NoStop}%
\bibitem [{\citenamefont {Giustina}\ \emph {et~al.}(2015)\citenamefont {Giustina}, \citenamefont {Versteegh}, \citenamefont {Wengerowsky}, \citenamefont {Handsteiner}, \citenamefont {Hochrainer}, \citenamefont {Phelan}, \citenamefont {Steinlechner}, \citenamefont {Kofler}, \citenamefont {Larsson}, \citenamefont {Abell\'an}, \citenamefont {Amaya}, \citenamefont {Pruneri}, \citenamefont {Mitchell}, \citenamefont {Beyer}, \citenamefont {Gerrits}, \citenamefont {Lita}, \citenamefont {Shalm}, \citenamefont {Nam}, \citenamefont {Scheidl}, \citenamefont {Ursin}, \citenamefont {Wittmann},\ and\ \citenamefont {Zeilinger}}]{GiustinaPRL2015}%
  \BibitemOpen
  \bibfield  {author} {\bibinfo {author} {\bibfnamefont {Marissa}\ \bibnamefont {Giustina}}, \bibinfo {author} {\bibfnamefont {Marijn A.~M.}\ \bibnamefont {Versteegh}}, \bibinfo {author} {\bibfnamefont {S\"oren}\ \bibnamefont {Wengerowsky}}, \bibinfo {author} {\bibfnamefont {Johannes}\ \bibnamefont {Handsteiner}}, \bibinfo {author} {\bibfnamefont {Armin}\ \bibnamefont {Hochrainer}}, \bibinfo {author} {\bibfnamefont {Kevin}\ \bibnamefont {Phelan}}, \bibinfo {author} {\bibfnamefont {Fabian}\ \bibnamefont {Steinlechner}}, \bibinfo {author} {\bibfnamefont {Johannes}\ \bibnamefont {Kofler}}, \bibinfo {author} {\bibfnamefont {Jan-\AA{}ke}\ \bibnamefont {Larsson}}, \bibinfo {author} {\bibfnamefont {Carlos}\ \bibnamefont {Abell\'an}}, \bibinfo {author} {\bibfnamefont {Waldimar}\ \bibnamefont {Amaya}}, \bibinfo {author} {\bibfnamefont {Valerio}\ \bibnamefont {Pruneri}}, \bibinfo {author} {\bibfnamefont {Morgan~W.}\ \bibnamefont {Mitchell}}, \bibinfo {author} {\bibfnamefont {J\"orn}\ \bibnamefont {Beyer}}, \bibinfo
  {author} {\bibfnamefont {Thomas}\ \bibnamefont {Gerrits}}, \bibinfo {author} {\bibfnamefont {Adriana~E.}\ \bibnamefont {Lita}}, \bibinfo {author} {\bibfnamefont {Lynden~K.}\ \bibnamefont {Shalm}}, \bibinfo {author} {\bibfnamefont {Sae~Woo}\ \bibnamefont {Nam}}, \bibinfo {author} {\bibfnamefont {Thomas}\ \bibnamefont {Scheidl}}, \bibinfo {author} {\bibfnamefont {Rupert}\ \bibnamefont {Ursin}}, \bibinfo {author} {\bibfnamefont {Bernhard}\ \bibnamefont {Wittmann}}, \ and\ \bibinfo {author} {\bibfnamefont {Anton}\ \bibnamefont {Zeilinger}},\ }\bibfield  {title} {\enquote {\bibinfo {title} {Significant-loophole-free test of {B}ell's theorem with entangled photons},}\ }\href {\doibase 10.1103/PhysRevLett.115.250401} {\bibfield  {journal} {\bibinfo  {journal} {Physical Review Letters}\ }\textbf {\bibinfo {volume} {115}},\ \bibinfo {pages} {250401} (\bibinfo {year} {2015})}\BibitemShut {NoStop}%
\bibitem [{\citenamefont {Shalm}\ \emph {et~al.}(2015)\citenamefont {Shalm}, \citenamefont {Meyer-Scott}, \citenamefont {Christensen}, \citenamefont {Bierhorst}, \citenamefont {Wayne}, \citenamefont {Stevens}, \citenamefont {Gerrits}, \citenamefont {Glancy}, \citenamefont {Hamel}, \citenamefont {Allman}, \citenamefont {Coakley}, \citenamefont {Dyer}, \citenamefont {Hodge}, \citenamefont {Lita}, \citenamefont {Verma}, \citenamefont {Lambrocco}, \citenamefont {Tortorici}, \citenamefont {Migdall}, \citenamefont {Zhang}, \citenamefont {Kumor}, \citenamefont {Farr}, \citenamefont {Marsili}, \citenamefont {Shaw}, \citenamefont {Stern}, \citenamefont {Abell\'an}, \citenamefont {Amaya}, \citenamefont {Pruneri}, \citenamefont {Jennewein}, \citenamefont {Mitchell}, \citenamefont {Kwiat}, \citenamefont {Bienfang}, \citenamefont {Mirin}, \citenamefont {Knill},\ and\ \citenamefont {Nam}}]{ShalmPRL2015}%
  \BibitemOpen
  \bibfield  {author} {\bibinfo {author} {\bibfnamefont {Lynden~K.}\ \bibnamefont {Shalm}}, \bibinfo {author} {\bibfnamefont {Evan}\ \bibnamefont {Meyer-Scott}}, \bibinfo {author} {\bibfnamefont {Bradley~G.}\ \bibnamefont {Christensen}}, \bibinfo {author} {\bibfnamefont {Peter}\ \bibnamefont {Bierhorst}}, \bibinfo {author} {\bibfnamefont {Michael~A.}\ \bibnamefont {Wayne}}, \bibinfo {author} {\bibfnamefont {Martin~J.}\ \bibnamefont {Stevens}}, \bibinfo {author} {\bibfnamefont {Thomas}\ \bibnamefont {Gerrits}}, \bibinfo {author} {\bibfnamefont {Scott}\ \bibnamefont {Glancy}}, \bibinfo {author} {\bibfnamefont {Deny~R.}\ \bibnamefont {Hamel}}, \bibinfo {author} {\bibfnamefont {Michael~S.}\ \bibnamefont {Allman}}, \bibinfo {author} {\bibfnamefont {Kevin~J.}\ \bibnamefont {Coakley}}, \bibinfo {author} {\bibfnamefont {Shellee~D.}\ \bibnamefont {Dyer}}, \bibinfo {author} {\bibfnamefont {Carson}\ \bibnamefont {Hodge}}, \bibinfo {author} {\bibfnamefont {Adriana~E.}\ \bibnamefont {Lita}}, \bibinfo {author}
  {\bibfnamefont {Varun~B.}\ \bibnamefont {Verma}}, \bibinfo {author} {\bibfnamefont {Camilla}\ \bibnamefont {Lambrocco}}, \bibinfo {author} {\bibfnamefont {Edward}\ \bibnamefont {Tortorici}}, \bibinfo {author} {\bibfnamefont {Alan~L.}\ \bibnamefont {Migdall}}, \bibinfo {author} {\bibfnamefont {Yanbao}\ \bibnamefont {Zhang}}, \bibinfo {author} {\bibfnamefont {Daniel~R.}\ \bibnamefont {Kumor}}, \bibinfo {author} {\bibfnamefont {William~H.}\ \bibnamefont {Farr}}, \bibinfo {author} {\bibfnamefont {Francesco}\ \bibnamefont {Marsili}}, \bibinfo {author} {\bibfnamefont {Matthew~D.}\ \bibnamefont {Shaw}}, \bibinfo {author} {\bibfnamefont {Jeffrey~A.}\ \bibnamefont {Stern}}, \bibinfo {author} {\bibfnamefont {Carlos}\ \bibnamefont {Abell\'an}}, \bibinfo {author} {\bibfnamefont {Waldimar}\ \bibnamefont {Amaya}}, \bibinfo {author} {\bibfnamefont {Valerio}\ \bibnamefont {Pruneri}}, \bibinfo {author} {\bibfnamefont {Thomas}\ \bibnamefont {Jennewein}}, \bibinfo {author} {\bibfnamefont {Morgan~W.}\ \bibnamefont {Mitchell}},
  \bibinfo {author} {\bibfnamefont {Paul~G.}\ \bibnamefont {Kwiat}}, \bibinfo {author} {\bibfnamefont {Joshua~C.}\ \bibnamefont {Bienfang}}, \bibinfo {author} {\bibfnamefont {Richard~P.}\ \bibnamefont {Mirin}}, \bibinfo {author} {\bibfnamefont {Emanuel}\ \bibnamefont {Knill}}, \ and\ \bibinfo {author} {\bibfnamefont {Sae~Woo}\ \bibnamefont {Nam}},\ }\bibfield  {title} {\enquote {\bibinfo {title} {Strong loophole-free test of local realism},}\ }\href {\doibase 10.1103/PhysRevLett.115.250402} {\bibfield  {journal} {\bibinfo  {journal} {Physical Review Letters}\ }\textbf {\bibinfo {volume} {115}},\ \bibinfo {pages} {250402} (\bibinfo {year} {2015})}\BibitemShut {NoStop}%
\bibitem [{\citenamefont {Zhong}\ \emph {et~al.}(2020)\citenamefont {Zhong}, \citenamefont {Wang}, \citenamefont {Deng}, \citenamefont {Chen}, \citenamefont {Peng}, \citenamefont {Luo}, \citenamefont {Qin}, \citenamefont {Wu}, \citenamefont {Ding}, \citenamefont {Hu}, \citenamefont {Hu}, \citenamefont {Yang}, \citenamefont {Zhang}, \citenamefont {Li}, \citenamefont {Li}, \citenamefont {Jiang}, \citenamefont {Gan}, \citenamefont {Yang}, \citenamefont {You}, \citenamefont {Wang}, \citenamefont {Li}, \citenamefont {Liu}, \citenamefont {Lu},\ and\ \citenamefont {Pan}}]{ZhongS2020}%
  \BibitemOpen
  \bibfield  {author} {\bibinfo {author} {\bibfnamefont {Han-Sen}\ \bibnamefont {Zhong}}, \bibinfo {author} {\bibfnamefont {Hui}\ \bibnamefont {Wang}}, \bibinfo {author} {\bibfnamefont {Yu-Hao}\ \bibnamefont {Deng}}, \bibinfo {author} {\bibfnamefont {Ming-Cheng}\ \bibnamefont {Chen}}, \bibinfo {author} {\bibfnamefont {Li-Chao}\ \bibnamefont {Peng}}, \bibinfo {author} {\bibfnamefont {Yi-Han}\ \bibnamefont {Luo}}, \bibinfo {author} {\bibfnamefont {Jian}\ \bibnamefont {Qin}}, \bibinfo {author} {\bibfnamefont {Dian}\ \bibnamefont {Wu}}, \bibinfo {author} {\bibfnamefont {Xing}\ \bibnamefont {Ding}}, \bibinfo {author} {\bibfnamefont {Yi}~\bibnamefont {Hu}}, \bibinfo {author} {\bibfnamefont {Peng}\ \bibnamefont {Hu}}, \bibinfo {author} {\bibfnamefont {Xiao-Yan}\ \bibnamefont {Yang}}, \bibinfo {author} {\bibfnamefont {Wei-Jun}\ \bibnamefont {Zhang}}, \bibinfo {author} {\bibfnamefont {Hao}\ \bibnamefont {Li}}, \bibinfo {author} {\bibfnamefont {Yuxuan}\ \bibnamefont {Li}}, \bibinfo {author} {\bibfnamefont {Xiao}\
  \bibnamefont {Jiang}}, \bibinfo {author} {\bibfnamefont {Lin}\ \bibnamefont {Gan}}, \bibinfo {author} {\bibfnamefont {Guangwen}\ \bibnamefont {Yang}}, \bibinfo {author} {\bibfnamefont {Lixing}\ \bibnamefont {You}}, \bibinfo {author} {\bibfnamefont {Zhen}\ \bibnamefont {Wang}}, \bibinfo {author} {\bibfnamefont {Li}~\bibnamefont {Li}}, \bibinfo {author} {\bibfnamefont {Nai-Le}\ \bibnamefont {Liu}}, \bibinfo {author} {\bibfnamefont {Chao-Yang}\ \bibnamefont {Lu}}, \ and\ \bibinfo {author} {\bibfnamefont {Jian-Wei}\ \bibnamefont {Pan}},\ }\bibfield  {title} {\enquote {\bibinfo {title} {Quantum computational advantage using photons},}\ }\href {\doibase 10.1126/science.abe8770} {\bibfield  {journal} {\bibinfo  {journal} {Science}\ }\textbf {\bibinfo {volume} {370}},\ \bibinfo {pages} {1460--1463} (\bibinfo {year} {2020})}\BibitemShut {NoStop}%
\bibitem [{\citenamefont {Becker-Hickl}(2023)}]{PMT-BeckerHickl}%
  \BibitemOpen
  \bibfield  {author} {\bibinfo {author} {\bibnamefont {Becker-Hickl}},\ }\bibfield  {title} {\enquote {\bibinfo {title} {Cooled fast {PMT} modules: {PMC-150}},}\ }\href {https://www.becker-hickl.com/products/cooled-fast-pmt-modules/} {\bibfield  {journal} {\bibinfo  {journal} {Datasheet db-pmc150-v02}\ } (\bibinfo {year} {2023})}\BibitemShut {NoStop}%
\bibitem [{EMC(2024)}]{EMCCD-iKon}%
  \BibitemOpen
  \bibfield  {title} {\enquote {\bibinfo {title} {{Oxford Instruments}: {iKon} slow-scan large {CCD} cameras: {iKon-XL 231}},}\ }\href {https://andor.oxinst.com/products/ikon-slow-scan-cameras-for-physical-science} {\bibfield  {journal} {\bibinfo  {journal} {Datasheet iKonXL231SS 0622 R1}\ } (\bibinfo {year} {2024})}\BibitemShut {NoStop}%
\bibitem [{CMO(2023)}]{CMOS-Retiga}%
  \BibitemOpen
  \bibfield  {title} {\enquote {\bibinfo {title} {{Teledyne Technologies}: Long exposure optimized {CMOS} camera: Retiga {E20}},}\ }\href {https://www.photometrics.com/wp-content/uploads/2022/10/Teledyne-Retiga-E7-Camera.pdf} {\bibfield  {journal} {\bibinfo  {journal} {Retiga E-Series CMOS Camera Datasheet Rev A0-01-28-2023}\ } (\bibinfo {year} {2023})}\BibitemShut {NoStop}%
\bibitem [{\citenamefont {Hamamatsu}(2023)}]{CMOS-ORCA}%
  \BibitemOpen
  \bibfield  {author} {\bibinfo {author} {\bibnamefont {Hamamatsu}},\ }\bibfield  {title} {\enquote {\bibinfo {title} {{ORCA-Quest} {qCMOS} camera: {C15550-20UP}},}\ }\href {https://www.hamamatsu.com/content/dam/hamamatsu-photonics/sites/documents/99_SALES_LIBRARY/sys/SCAS0154E_C15550-20UP_tec.pdf} {\bibfield  {journal} {\bibinfo  {journal} {Technical note Cat. No. SCAS0154E05 NOV/2023 HPK}\ } (\bibinfo {year} {2023})}\BibitemShut {NoStop}%
\bibitem [{\citenamefont {Dreyling-Eschweiler}\ \emph {et~al.}(2015)\citenamefont {Dreyling-Eschweiler}, \citenamefont {Bastidon}, \citenamefont {Döbrich}, \citenamefont {Horns}, \citenamefont {Januschek},\ and\ \citenamefont {Lindner}}]{Dreyling-Eschweiler2015}%
  \BibitemOpen
  \bibfield  {author} {\bibinfo {author} {\bibfnamefont {J}~\bibnamefont {Dreyling-Eschweiler}}, \bibinfo {author} {\bibfnamefont {N.}~\bibnamefont {Bastidon}}, \bibinfo {author} {\bibfnamefont {B.}~\bibnamefont {Döbrich}}, \bibinfo {author} {\bibfnamefont {D.}~\bibnamefont {Horns}}, \bibinfo {author} {\bibfnamefont {F.}~\bibnamefont {Januschek}}, \ and\ \bibinfo {author} {\bibfnamefont {A.}~\bibnamefont {Lindner}},\ }\bibfield  {title} {\enquote {\bibinfo {title} {Characterization, 1064 nm photon signals and background events of a tungsten {TES} detector for the {ALPS} experiment},}\ }\href {https://doi.org/10.1080/09500340.2015.1021723} {\bibfield  {journal} {\bibinfo  {journal} {Journal of Modern Optics}\ } (\bibinfo {year} {2015})}\BibitemShut {NoStop}%
\bibitem [{SNS(2024)}]{SNSPD-SingleQuantum}%
  \BibitemOpen
  \bibfield  {title} {\enquote {\bibinfo {title} {{Single Quantum Eos}: {SNSPD} closed-cycle system},}\ }\href {https://www.singlequantum.com/wp-content/uploads/2022/12/SQ-General-Brochure.pdf} {\bibfield  {journal} {\bibinfo  {journal} {Datasheet SQ A4 brochure v6}\ } (\bibinfo {year} {2024})}\BibitemShut {NoStop}%
\bibitem [{\citenamefont {Sperlich}\ and\ \citenamefont {Stolz}(2013)}]{SperlichMST2013}%
  \BibitemOpen
  \bibfield  {author} {\bibinfo {author} {\bibfnamefont {Karsten}\ \bibnamefont {Sperlich}}\ and\ \bibinfo {author} {\bibfnamefont {Heinrich}\ \bibnamefont {Stolz}},\ }\bibfield  {title} {\enquote {\bibinfo {title} {Quantum efficiency measurements of {(EM)CCD} cameras: high spectral resolution and temperature dependence},}\ }\href {\doibase 10.1088/0957-0233/25/1/015502} {\bibfield  {journal} {\bibinfo  {journal} {Measurement Science and Technology}\ }\textbf {\bibinfo {volume} {25}},\ \bibinfo {pages} {015502} (\bibinfo {year} {2013})}\BibitemShut {NoStop}%
\bibitem [{\citenamefont {Shao}\ \emph {et~al.}(2021)\citenamefont {Shao}, \citenamefont {Cheng}, \citenamefont {Xu},\ and\ \citenamefont {Song}}]{ShaoOQE2021}%
  \BibitemOpen
  \bibfield  {author} {\bibinfo {author} {\bibfnamefont {Hanxiao}\ \bibnamefont {Shao}}, \bibinfo {author} {\bibfnamefont {Bo}~\bibnamefont {Cheng}}, \bibinfo {author} {\bibfnamefont {Yun}\ \bibnamefont {Xu}}, \ and\ \bibinfo {author} {\bibfnamefont {Guofeng}\ \bibnamefont {Song}},\ }\bibfield  {title} {\enquote {\bibinfo {title} {Ultrahigh-quantum-efficiency and high-bandwidth nanowire array {UTC-PD}s working at 1064 nm},}\ }\href {\doibase 10.1007/s11082-021-03293-0} {\bibfield  {journal} {\bibinfo  {journal} {Optical and Quantum Electronics}\ }\textbf {\bibinfo {volume} {54}},\ \bibinfo {pages} {15} (\bibinfo {year} {2021})}\BibitemShut {NoStop}%
\bibitem [{\citenamefont {Hochberg}\ \emph {et~al.}(2019)\citenamefont {Hochberg}, \citenamefont {Charaev}, \citenamefont {Nam}, \citenamefont {Verma}, \citenamefont {Colangelo},\ and\ \citenamefont {Berggren}}]{HochbergPRL2019}%
  \BibitemOpen
  \bibfield  {author} {\bibinfo {author} {\bibfnamefont {Yonit}\ \bibnamefont {Hochberg}}, \bibinfo {author} {\bibfnamefont {Ilya}\ \bibnamefont {Charaev}}, \bibinfo {author} {\bibfnamefont {Sae-Woo}\ \bibnamefont {Nam}}, \bibinfo {author} {\bibfnamefont {Varun}\ \bibnamefont {Verma}}, \bibinfo {author} {\bibfnamefont {Marco}\ \bibnamefont {Colangelo}}, \ and\ \bibinfo {author} {\bibfnamefont {Karl~K.}\ \bibnamefont {Berggren}},\ }\bibfield  {title} {\enquote {\bibinfo {title} {Detecting sub-{GeV} dark matter with superconducting nanowires},}\ }\href {\doibase 10.1103/PhysRevLett.123.151802} {\bibfield  {journal} {\bibinfo  {journal} {Phys. Rev. Lett.}\ }\textbf {\bibinfo {volume} {123}},\ \bibinfo {pages} {151802} (\bibinfo {year} {2019})}\BibitemShut {NoStop}%
\bibitem [{\citenamefont {Lo}\ \emph {et~al.}(2005)\citenamefont {Lo}, \citenamefont {Ma},\ and\ \citenamefont {Chen}}]{LoPRL2004}%
  \BibitemOpen
  \bibfield  {author} {\bibinfo {author} {\bibfnamefont {Hoi-Kwong}\ \bibnamefont {Lo}}, \bibinfo {author} {\bibfnamefont {Xiongfeng}\ \bibnamefont {Ma}}, \ and\ \bibinfo {author} {\bibfnamefont {Kai}\ \bibnamefont {Chen}},\ }\bibfield  {title} {\enquote {\bibinfo {title} {Decoy state quantum key distribution},}\ }\href {\doibase 10.1103/PhysRevLett.94.230504} {\bibfield  {journal} {\bibinfo  {journal} {Phys. Rev. Lett.}\ }\textbf {\bibinfo {volume} {94}},\ \bibinfo {pages} {230504} (\bibinfo {year} {2005})}\BibitemShut {NoStop}%
\bibitem [{\citenamefont {Liao}\ \emph {et~al.}(2017)\citenamefont {Liao}, \citenamefont {Yong}, \citenamefont {Liu}, \citenamefont {Shentu}, \citenamefont {Li}, \citenamefont {Lin}, \citenamefont {Dai}, \citenamefont {Zhao}, \citenamefont {Li}, \citenamefont {Guan}, \citenamefont {Chen}, \citenamefont {Gong}, \citenamefont {Li}, \citenamefont {Lin}, \citenamefont {Pan}, \citenamefont {Pelc}, \citenamefont {Fejer}, \citenamefont {Zhang}, \citenamefont {Liu}, \citenamefont {Yin}, \citenamefont {Ren}, \citenamefont {Wang}, \citenamefont {Zhang}, \citenamefont {Peng},\ and\ \citenamefont {Pan}}]{Liao2017}%
  \BibitemOpen
  \bibfield  {author} {\bibinfo {author} {\bibfnamefont {SK.}\ \bibnamefont {Liao}}, \bibinfo {author} {\bibfnamefont {HL.}\ \bibnamefont {Yong}}, \bibinfo {author} {\bibfnamefont {C.}~\bibnamefont {Liu}}, \bibinfo {author} {\bibfnamefont {GL.}\ \bibnamefont {Shentu}}, \bibinfo {author} {\bibfnamefont {DD.}\ \bibnamefont {Li}}, \bibinfo {author} {\bibfnamefont {J.}~\bibnamefont {Lin}}, \bibinfo {author} {\bibfnamefont {H.}~\bibnamefont {Dai}}, \bibinfo {author} {\bibfnamefont {SQ.}\ \bibnamefont {Zhao}}, \bibinfo {author} {\bibfnamefont {B.}~\bibnamefont {Li}}, \bibinfo {author} {\bibfnamefont {JY.}\ \bibnamefont {Guan}}, \bibinfo {author} {\bibfnamefont {W.}~\bibnamefont {Chen}}, \bibinfo {author} {\bibfnamefont {YH.}\ \bibnamefont {Gong}}, \bibinfo {author} {\bibfnamefont {Y.}~\bibnamefont {Li}}, \bibinfo {author} {\bibfnamefont {ZH.}\ \bibnamefont {Lin}}, \bibinfo {author} {\bibfnamefont {GS.}\ \bibnamefont {Pan}}, \bibinfo {author} {\bibfnamefont {J.S.}\ \bibnamefont {Pelc}}, \bibinfo {author}
  {\bibfnamefont {M.M.}\ \bibnamefont {Fejer}}, \bibinfo {author} {\bibfnamefont {WZ.}\ \bibnamefont {Zhang}}, \bibinfo {author} {\bibfnamefont {WY.}\ \bibnamefont {Liu}}, \bibinfo {author} {\bibfnamefont {J.}~\bibnamefont {Yin}}, \bibinfo {author} {\bibfnamefont {JG.}\ \bibnamefont {Ren}}, \bibinfo {author} {\bibfnamefont {XB.}\ \bibnamefont {Wang}}, \bibinfo {author} {\bibfnamefont {Q.}~\bibnamefont {Zhang}}, \bibinfo {author} {\bibfnamefont {CZ.}\ \bibnamefont {Peng}}, \ and\ \bibinfo {author} {\bibfnamefont {JW.}\ \bibnamefont {Pan}},\ }\bibfield  {title} {\enquote {\bibinfo {title} {Long-distance free-space quantum key distribution in daylight towards inter-satellite communication},}\ }\href {https://doi.org/10.1038/nphoton.2017.116} {\bibfield  {journal} {\bibinfo  {journal} {Nature Photonics}\ } (\bibinfo {year} {2017})}\BibitemShut {NoStop}%
\bibitem [{\citenamefont {Klop}\ \emph {et~al.}(2021)\citenamefont {Klop}, \citenamefont {Saathof}, \citenamefont {Doelman}, \citenamefont {Gruber}, \citenamefont {Moens}, \citenamefont {Osorio~Tamayo},\ and\ \citenamefont {Duque}}]{Klop2021}%
  \BibitemOpen
  \bibfield  {author} {\bibinfo {author} {\bibfnamefont {W.}~\bibnamefont {Klop}}, \bibinfo {author} {\bibfnamefont {R.}~\bibnamefont {Saathof}}, \bibinfo {author} {\bibfnamefont {N.}~\bibnamefont {Doelman}}, \bibinfo {author} {\bibfnamefont {M.}~\bibnamefont {Gruber}}, \bibinfo {author} {\bibfnamefont {T.}~\bibnamefont {Moens}}, \bibinfo {author} {\bibfnamefont {C.I.}\ \bibnamefont {Osorio~Tamayo}}, \ and\ \bibinfo {author} {\bibfnamefont {C.}~\bibnamefont {Duque}},\ }\bibfield  {title} {\enquote {\bibinfo {title} {Qkd optical ground terminal developments},}\ }\href {https://doi.org/10.1117/12.2599217} {\bibfield  {journal} {\bibinfo  {journal} {International Conference on Space Optics Proceedings}\ } (\bibinfo {year} {2021})}\BibitemShut {NoStop}%
\bibitem [{\citenamefont {Yang}\ \emph {et~al.}(2020)\citenamefont {Yang}, \citenamefont {Abulizi}, \citenamefont {Li}, \citenamefont {Zhang}, \citenamefont {Li}, \citenamefont {Liu}, \citenamefont {Yin}, \citenamefont {Cao}, \citenamefont {Ren},\ and\ \citenamefont {Peng}}]{Yang2020}%
  \BibitemOpen
  \bibfield  {author} {\bibinfo {author} {\bibfnamefont {KX.}\ \bibnamefont {Yang}}, \bibinfo {author} {\bibfnamefont {M.}~\bibnamefont {Abulizi}}, \bibinfo {author} {\bibfnamefont {YH.}\ \bibnamefont {Li}}, \bibinfo {author} {\bibfnamefont {BY.}\ \bibnamefont {Zhang}}, \bibinfo {author} {\bibfnamefont {SL.}\ \bibnamefont {Li}}, \bibinfo {author} {\bibfnamefont {WY.}\ \bibnamefont {Liu}}, \bibinfo {author} {\bibfnamefont {J.}~\bibnamefont {Yin}}, \bibinfo {author} {\bibfnamefont {Y.}~\bibnamefont {Cao}}, \bibinfo {author} {\bibfnamefont {JG.}\ \bibnamefont {Ren}}, \ and\ \bibinfo {author} {\bibfnamefont {CZ.}\ \bibnamefont {Peng}},\ }\bibfield  {title} {\enquote {\bibinfo {title} {Single-mode fiber coupling with a m-spgd algorithm for long-range quantum communications},}\ }\href {https://doi.org/10.1364/OE.411939} {\bibfield  {journal} {\bibinfo  {journal} {Optics Express}\ } (\bibinfo {year} {2020})}\BibitemShut {NoStop}%
\bibitem [{\citenamefont {Chiles}\ \emph {et~al.}(2022)\citenamefont {Chiles}, \citenamefont {Charaev}, \citenamefont {Lasenby}, \citenamefont {Baryakhtar}, \citenamefont {Huang}, \citenamefont {Roshko}, \citenamefont {Burton}, \citenamefont {Colangelo}, \citenamefont {Van~Tilburg}, \citenamefont {Arvanitaki}, \citenamefont {Nam},\ and\ \citenamefont {Berggren}}]{ChilesPRL2022}%
  \BibitemOpen
  \bibfield  {author} {\bibinfo {author} {\bibfnamefont {Jeff}\ \bibnamefont {Chiles}}, \bibinfo {author} {\bibfnamefont {Ilya}\ \bibnamefont {Charaev}}, \bibinfo {author} {\bibfnamefont {Robert}\ \bibnamefont {Lasenby}}, \bibinfo {author} {\bibfnamefont {Masha}\ \bibnamefont {Baryakhtar}}, \bibinfo {author} {\bibfnamefont {Junwu}\ \bibnamefont {Huang}}, \bibinfo {author} {\bibfnamefont {Alexana}\ \bibnamefont {Roshko}}, \bibinfo {author} {\bibfnamefont {George}\ \bibnamefont {Burton}}, \bibinfo {author} {\bibfnamefont {Marco}\ \bibnamefont {Colangelo}}, \bibinfo {author} {\bibfnamefont {Ken}\ \bibnamefont {Van~Tilburg}}, \bibinfo {author} {\bibfnamefont {Asimina}\ \bibnamefont {Arvanitaki}}, \bibinfo {author} {\bibfnamefont {Sae~Woo}\ \bibnamefont {Nam}}, \ and\ \bibinfo {author} {\bibfnamefont {Karl~K.}\ \bibnamefont {Berggren}},\ }\bibfield  {title} {\enquote {\bibinfo {title} {New constraints on dark photon dark matter with superconducting nanowire detectors in an optical haloscope},}\ }\href {\doibase
  10.1103/PhysRevLett.128.231802} {\bibfield  {journal} {\bibinfo  {journal} {Phys. Rev. Lett.}\ }\textbf {\bibinfo {volume} {128}},\ \bibinfo {pages} {231802} (\bibinfo {year} {2022})}\BibitemShut {NoStop}%
\bibitem [{\citenamefont {Spector}(2023)}]{SpectorInKimballBook2023}%
  \BibitemOpen
  \bibfield  {author} {\bibinfo {author} {\bibfnamefont {Aaron~D.}\ \bibnamefont {Spector}},\ }\enquote {\bibinfo {title} {Light-shining-through-walls experiments},}\ in\ \href {\doibase 10.1007/978-3-030-95852-7_9} {\emph {\bibinfo {booktitle} {The Search for Ultralight Bosonic Dark Matter}}},\ \bibinfo {editor} {edited by\ \bibinfo {editor} {\bibfnamefont {Derek~F.}\ \bibnamefont {Jackson~Kimball}}\ and\ \bibinfo {editor} {\bibfnamefont {Karl}\ \bibnamefont {van Bibber}}}\ (\bibinfo  {publisher} {Springer International Publishing},\ \bibinfo {address} {Cham},\ \bibinfo {year} {2023})\ pp.\ \bibinfo {pages} {255--279}\BibitemShut {NoStop}%
\bibitem [{\citenamefont {Chen}(2022)}]{ChenOQE2022}%
  \BibitemOpen
  \bibfield  {author} {\bibinfo {author} {\bibfnamefont {Shun-Ping}\ \bibnamefont {Chen}},\ }\bibfield  {title} {\enquote {\bibinfo {title} {Performance analysis of near-earth, lunar and interplanetary optical communication links},}\ }\href {\doibase 10.1007/s11082-022-03987-z} {\bibfield  {journal} {\bibinfo  {journal} {Optical and Quantum Electronics}\ }\textbf {\bibinfo {volume} {54}},\ \bibinfo {pages} {562} (\bibinfo {year} {2022})}\BibitemShut {NoStop}%
\bibitem [{\citenamefont {Rustamkulov}\ \emph {et~al.}(2023)\citenamefont {Rustamkulov}, \citenamefont {Sing}, \citenamefont {Mukherjee}, \citenamefont {May}, \citenamefont {Kirk}, \citenamefont {Schlawin}, \citenamefont {Line}, \citenamefont {Piaulet}, \citenamefont {Carter}, \citenamefont {Batalha}, \citenamefont {Goyal}, \citenamefont {L{\'o}pez-Morales}, \citenamefont {Lothringer}, \citenamefont {MacDonald}, \citenamefont {Moran}, \citenamefont {Stevenson}, \citenamefont {Wakeford}, \citenamefont {Espinoza}, \citenamefont {Bean}, \citenamefont {Batalha}, \citenamefont {Benneke}, \citenamefont {Berta-Thompson}, \citenamefont {Crossfield}, \citenamefont {Gao}, \citenamefont {Kreidberg}, \citenamefont {Powell}, \citenamefont {Cubillos}, \citenamefont {Gibson}, \citenamefont {Leconte}, \citenamefont {Molaverdikhani}, \citenamefont {Nikolov}, \citenamefont {Parmentier}, \citenamefont {Roy}, \citenamefont {Taylor}, \citenamefont {Turner}, \citenamefont {Wheatley}, \citenamefont {Aggarwal}, \citenamefont
  {Ahrer}, \citenamefont {Alam}, \citenamefont {Alderson}, \citenamefont {Allen}, \citenamefont {Banerjee}, \citenamefont {Barat}, \citenamefont {Barrado}, \citenamefont {Barstow}, \citenamefont {Bell}, \citenamefont {Blecic}, \citenamefont {Brande}, \citenamefont {Casewell}, \citenamefont {Changeat}, \citenamefont {Chubb}, \citenamefont {Crouzet}, \citenamefont {Daylan}, \citenamefont {Decin}, \citenamefont {D{\'e}sert}, \citenamefont {Mikal-Evans}, \citenamefont {Feinstein}, \citenamefont {Flagg}, \citenamefont {Fortney}, \citenamefont {Harrington}, \citenamefont {Heng}, \citenamefont {Hong}, \citenamefont {Hu}, \citenamefont {Iro}, \citenamefont {Kataria}, \citenamefont {Kempton}, \citenamefont {Krick}, \citenamefont {Lendl}, \citenamefont {Lillo-Box}, \citenamefont {Louca}, \citenamefont {Lustig-Yaeger}, \citenamefont {Mancini}, \citenamefont {Mansfield}, \citenamefont {Mayne}, \citenamefont {Miguel}, \citenamefont {Morello}, \citenamefont {Ohno}, \citenamefont {Palle}, \citenamefont {Petit dit de~la
  Roche}, \citenamefont {Rackham}, \citenamefont {Radica}, \citenamefont {Ramos-Rosado}, \citenamefont {Redfield}, \citenamefont {Rogers}, \citenamefont {Shkolnik}, \citenamefont {Southworth}, \citenamefont {Teske}, \citenamefont {Tremblin}, \citenamefont {Tucker}, \citenamefont {Venot}, \citenamefont {Waalkes}, \citenamefont {Welbanks}, \citenamefont {Zhang},\ and\ \citenamefont {Zieba}}]{RustamkulovN2023}%
  \BibitemOpen
  \bibfield  {author} {\bibinfo {author} {\bibfnamefont {Z.}~\bibnamefont {Rustamkulov}}, \bibinfo {author} {\bibfnamefont {D.~K.}\ \bibnamefont {Sing}}, \bibinfo {author} {\bibfnamefont {S.}~\bibnamefont {Mukherjee}}, \bibinfo {author} {\bibfnamefont {E.~M.}\ \bibnamefont {May}}, \bibinfo {author} {\bibfnamefont {J.}~\bibnamefont {Kirk}}, \bibinfo {author} {\bibfnamefont {E.}~\bibnamefont {Schlawin}}, \bibinfo {author} {\bibfnamefont {M.~R.}\ \bibnamefont {Line}}, \bibinfo {author} {\bibfnamefont {C.}~\bibnamefont {Piaulet}}, \bibinfo {author} {\bibfnamefont {A.~L.}\ \bibnamefont {Carter}}, \bibinfo {author} {\bibfnamefont {N.~E.}\ \bibnamefont {Batalha}}, \bibinfo {author} {\bibfnamefont {J.~M.}\ \bibnamefont {Goyal}}, \bibinfo {author} {\bibfnamefont {M.}~\bibnamefont {L{\'o}pez-Morales}}, \bibinfo {author} {\bibfnamefont {J.~D.}\ \bibnamefont {Lothringer}}, \bibinfo {author} {\bibfnamefont {R.~J.}\ \bibnamefont {MacDonald}}, \bibinfo {author} {\bibfnamefont {S.~E.}\ \bibnamefont {Moran}}, \bibinfo
  {author} {\bibfnamefont {K.~B.}\ \bibnamefont {Stevenson}}, \bibinfo {author} {\bibfnamefont {H.~R.}\ \bibnamefont {Wakeford}}, \bibinfo {author} {\bibfnamefont {N.}~\bibnamefont {Espinoza}}, \bibinfo {author} {\bibfnamefont {J.~L.}\ \bibnamefont {Bean}}, \bibinfo {author} {\bibfnamefont {N.~M.}\ \bibnamefont {Batalha}}, \bibinfo {author} {\bibfnamefont {B.}~\bibnamefont {Benneke}}, \bibinfo {author} {\bibfnamefont {Z.~K.}\ \bibnamefont {Berta-Thompson}}, \bibinfo {author} {\bibfnamefont {I.~J.~M.}\ \bibnamefont {Crossfield}}, \bibinfo {author} {\bibfnamefont {P.}~\bibnamefont {Gao}}, \bibinfo {author} {\bibfnamefont {L.}~\bibnamefont {Kreidberg}}, \bibinfo {author} {\bibfnamefont {D.~K.}\ \bibnamefont {Powell}}, \bibinfo {author} {\bibfnamefont {P.~E.}\ \bibnamefont {Cubillos}}, \bibinfo {author} {\bibfnamefont {N.~P.}\ \bibnamefont {Gibson}}, \bibinfo {author} {\bibfnamefont {J.}~\bibnamefont {Leconte}}, \bibinfo {author} {\bibfnamefont {K.}~\bibnamefont {Molaverdikhani}}, \bibinfo {author} {\bibfnamefont
  {N.~K.}\ \bibnamefont {Nikolov}}, \bibinfo {author} {\bibfnamefont {V.}~\bibnamefont {Parmentier}}, \bibinfo {author} {\bibfnamefont {P.}~\bibnamefont {Roy}}, \bibinfo {author} {\bibfnamefont {J.}~\bibnamefont {Taylor}}, \bibinfo {author} {\bibfnamefont {J.~D.}\ \bibnamefont {Turner}}, \bibinfo {author} {\bibfnamefont {P.~J.}\ \bibnamefont {Wheatley}}, \bibinfo {author} {\bibfnamefont {K.}~\bibnamefont {Aggarwal}}, \bibinfo {author} {\bibfnamefont {E.}~\bibnamefont {Ahrer}}, \bibinfo {author} {\bibfnamefont {M.~K.}\ \bibnamefont {Alam}}, \bibinfo {author} {\bibfnamefont {L.}~\bibnamefont {Alderson}}, \bibinfo {author} {\bibfnamefont {N.~H.}\ \bibnamefont {Allen}}, \bibinfo {author} {\bibfnamefont {A.}~\bibnamefont {Banerjee}}, \bibinfo {author} {\bibfnamefont {S.}~\bibnamefont {Barat}}, \bibinfo {author} {\bibfnamefont {D.}~\bibnamefont {Barrado}}, \bibinfo {author} {\bibfnamefont {J.~K.}\ \bibnamefont {Barstow}}, \bibinfo {author} {\bibfnamefont {T.~J.}\ \bibnamefont {Bell}}, \bibinfo {author}
  {\bibfnamefont {J.}~\bibnamefont {Blecic}}, \bibinfo {author} {\bibfnamefont {J.}~\bibnamefont {Brande}}, \bibinfo {author} {\bibfnamefont {S.}~\bibnamefont {Casewell}}, \bibinfo {author} {\bibfnamefont {Q.}~\bibnamefont {Changeat}}, \bibinfo {author} {\bibfnamefont {K.~L.}\ \bibnamefont {Chubb}}, \bibinfo {author} {\bibfnamefont {N.}~\bibnamefont {Crouzet}}, \bibinfo {author} {\bibfnamefont {T.}~\bibnamefont {Daylan}}, \bibinfo {author} {\bibfnamefont {L.}~\bibnamefont {Decin}}, \bibinfo {author} {\bibfnamefont {J.}~\bibnamefont {D{\'e}sert}}, \bibinfo {author} {\bibfnamefont {T.}~\bibnamefont {Mikal-Evans}}, \bibinfo {author} {\bibfnamefont {A.~D.}\ \bibnamefont {Feinstein}}, \bibinfo {author} {\bibfnamefont {L.}~\bibnamefont {Flagg}}, \bibinfo {author} {\bibfnamefont {J.~J.}\ \bibnamefont {Fortney}}, \bibinfo {author} {\bibfnamefont {J.}~\bibnamefont {Harrington}}, \bibinfo {author} {\bibfnamefont {K.}~\bibnamefont {Heng}}, \bibinfo {author} {\bibfnamefont {Y.}~\bibnamefont {Hong}}, \bibinfo {author}
  {\bibfnamefont {R.}~\bibnamefont {Hu}}, \bibinfo {author} {\bibfnamefont {N.}~\bibnamefont {Iro}}, \bibinfo {author} {\bibfnamefont {T.}~\bibnamefont {Kataria}}, \bibinfo {author} {\bibfnamefont {E.~M.~R.}\ \bibnamefont {Kempton}}, \bibinfo {author} {\bibfnamefont {J.}~\bibnamefont {Krick}}, \bibinfo {author} {\bibfnamefont {M.}~\bibnamefont {Lendl}}, \bibinfo {author} {\bibfnamefont {J.}~\bibnamefont {Lillo-Box}}, \bibinfo {author} {\bibfnamefont {A.}~\bibnamefont {Louca}}, \bibinfo {author} {\bibfnamefont {J.}~\bibnamefont {Lustig-Yaeger}}, \bibinfo {author} {\bibfnamefont {L.}~\bibnamefont {Mancini}}, \bibinfo {author} {\bibfnamefont {M.}~\bibnamefont {Mansfield}}, \bibinfo {author} {\bibfnamefont {N.~J.}\ \bibnamefont {Mayne}}, \bibinfo {author} {\bibfnamefont {Y.}~\bibnamefont {Miguel}}, \bibinfo {author} {\bibfnamefont {G.}~\bibnamefont {Morello}}, \bibinfo {author} {\bibfnamefont {K.}~\bibnamefont {Ohno}}, \bibinfo {author} {\bibfnamefont {E.}~\bibnamefont {Palle}}, \bibinfo {author} {\bibfnamefont
  {D.~J.~M.}\ \bibnamefont {Petit dit de~la Roche}}, \bibinfo {author} {\bibfnamefont {B.~V.}\ \bibnamefont {Rackham}}, \bibinfo {author} {\bibfnamefont {M.}~\bibnamefont {Radica}}, \bibinfo {author} {\bibfnamefont {L.}~\bibnamefont {Ramos-Rosado}}, \bibinfo {author} {\bibfnamefont {S.}~\bibnamefont {Redfield}}, \bibinfo {author} {\bibfnamefont {L.~K.}\ \bibnamefont {Rogers}}, \bibinfo {author} {\bibfnamefont {E.~L.}\ \bibnamefont {Shkolnik}}, \bibinfo {author} {\bibfnamefont {J.}~\bibnamefont {Southworth}}, \bibinfo {author} {\bibfnamefont {J.}~\bibnamefont {Teske}}, \bibinfo {author} {\bibfnamefont {P.}~\bibnamefont {Tremblin}}, \bibinfo {author} {\bibfnamefont {G.~S.}\ \bibnamefont {Tucker}}, \bibinfo {author} {\bibfnamefont {O.}~\bibnamefont {Venot}}, \bibinfo {author} {\bibfnamefont {W.~C.}\ \bibnamefont {Waalkes}}, \bibinfo {author} {\bibfnamefont {L.}~\bibnamefont {Welbanks}}, \bibinfo {author} {\bibfnamefont {X.}~\bibnamefont {Zhang}}, \ and\ \bibinfo {author} {\bibfnamefont {S.}~\bibnamefont
  {Zieba}},\ }\bibfield  {title} {\enquote {\bibinfo {title} {Early release science of the exoplanet {WASP-39b} with {JWST NIRSpec PRISM}},}\ }\href {\doibase 10.1038/s41586-022-05677-y} {\bibfield  {journal} {\bibinfo  {journal} {Nature}\ }\textbf {\bibinfo {volume} {614}},\ \bibinfo {pages} {659--663} (\bibinfo {year} {2023})}\BibitemShut {NoStop}%
\bibitem [{\citenamefont {Sliski}\ \emph {et~al.}(2023)\citenamefont {Sliski}, \citenamefont {Blake}, \citenamefont {Eastman},\ and\ \citenamefont {Halverson}}]{Sliski2023}%
  \BibitemOpen
  \bibfield  {author} {\bibinfo {author} {\bibfnamefont {D.H.}\ \bibnamefont {Sliski}}, \bibinfo {author} {\bibfnamefont {C.H.}\ \bibnamefont {Blake}}, \bibinfo {author} {\bibfnamefont {J.D.}\ \bibnamefont {Eastman}}, \ and\ \bibinfo {author} {\bibfnamefont {S.}~\bibnamefont {Halverson}},\ }\bibfield  {title} {\enquote {\bibinfo {title} {Seeing the limited coupling of starlight into single-mode fiber with a small telescope},}\ }\href {https://doi.org/10.1002/asna.20220080} {\bibfield  {journal} {\bibinfo  {journal} {Astronomical Nachrichten}\ } (\bibinfo {year} {2023})}\BibitemShut {NoStop}%
\bibitem [{\citenamefont {Crass}\ \emph {et~al.}(2020)\citenamefont {Crass}, \citenamefont {Bechter}, \citenamefont {Sands}, \citenamefont {King}, \citenamefont {Ketterer}, \citenamefont {Engstrom}, \citenamefont {Hamper}, \citenamefont {Kopon}, \citenamefont {Smous}, \citenamefont {Crepp}, \citenamefont {Montoya}, \citenamefont {O~Durney}, \citenamefont {Cavalieri}, \citenamefont {Reynolds}, \citenamefont {Vansickle}, \citenamefont {Onuma}, \citenamefont {Thomes}, \citenamefont {Mullin}, \citenamefont {Shelton}, \citenamefont {Wallace}, \citenamefont {Bechter}, \citenamefont {Vaz}, \citenamefont {Power}, \citenamefont {Rahmen},\ and\ \citenamefont {Ertel}}]{Crass2020}%
  \BibitemOpen
  \bibfield  {author} {\bibinfo {author} {\bibfnamefont {J.}~\bibnamefont {Crass}}, \bibinfo {author} {\bibfnamefont {A.}~\bibnamefont {Bechter}}, \bibinfo {author} {\bibfnamefont {B.}~\bibnamefont {Sands}}, \bibinfo {author} {\bibfnamefont {D.}~\bibnamefont {King}}, \bibinfo {author} {\bibfnamefont {R.}~\bibnamefont {Ketterer}}, \bibinfo {author} {\bibfnamefont {M.}~\bibnamefont {Engstrom}}, \bibinfo {author} {\bibfnamefont {R.}~\bibnamefont {Hamper}}, \bibinfo {author} {\bibfnamefont {D.}~\bibnamefont {Kopon}}, \bibinfo {author} {\bibfnamefont {J}~\bibnamefont {Smous}}, \bibinfo {author} {\bibfnamefont {J.R}\ \bibnamefont {Crepp}}, \bibinfo {author} {\bibfnamefont {M.}~\bibnamefont {Montoya}}, \bibinfo {author} {\bibfnamefont {O.}~\bibnamefont {O~Durney}}, \bibinfo {author} {\bibfnamefont {D.}~\bibnamefont {Cavalieri}}, \bibinfo {author} {\bibfnamefont {R.}~\bibnamefont {Reynolds}}, \bibinfo {author} {\bibfnamefont {M.}~\bibnamefont {Vansickle}}, \bibinfo {author} {\bibfnamefont {E.}~\bibnamefont {Onuma}},
  \bibinfo {author} {\bibfnamefont {J.}~\bibnamefont {Thomes}}, \bibinfo {author} {\bibfnamefont {S.}~\bibnamefont {Mullin}}, \bibinfo {author} {\bibfnamefont {C.}~\bibnamefont {Shelton}}, \bibinfo {author} {\bibfnamefont {K.}~\bibnamefont {Wallace}}, \bibinfo {author} {\bibfnamefont {E.}~\bibnamefont {Bechter}}, \bibinfo {author} {\bibfnamefont {A.}~\bibnamefont {Vaz}}, \bibinfo {author} {\bibfnamefont {J.}~\bibnamefont {Power}}, \bibinfo {author} {\bibfnamefont {G.}~\bibnamefont {Rahmen}}, \ and\ \bibinfo {author} {\bibfnamefont {S.}~\bibnamefont {Ertel}},\ }\bibfield  {title} {\enquote {\bibinfo {title} {Final design and on-sky testing of the ilocater sx acquisition camera: broad-band single-mode fibre coupling},}\ }\href {https://doi.org/10.1093/mnras/staa3355} {\bibfield  {journal} {\bibinfo  {journal} {Monthly Notices of the Royal Astronomical Society}\ } (\bibinfo {year} {2020})}\BibitemShut {NoStop}%
\bibitem [{\citenamefont {Zieli\'{n}ska}\ \emph {et~al.}(2012)\citenamefont {Zieli\'{n}ska}, \citenamefont {Beduini}, \citenamefont {Godbout},\ and\ \citenamefont {Mitchell}}]{ZielinskaOL2012}%
  \BibitemOpen
  \bibfield  {author} {\bibinfo {author} {\bibfnamefont {Joanna~A.}\ \bibnamefont {Zieli\'{n}ska}}, \bibinfo {author} {\bibfnamefont {Federica~A.}\ \bibnamefont {Beduini}}, \bibinfo {author} {\bibfnamefont {Nicolas}\ \bibnamefont {Godbout}}, \ and\ \bibinfo {author} {\bibfnamefont {Morgan~W.}\ \bibnamefont {Mitchell}},\ }\bibfield  {title} {\enquote {\bibinfo {title} {Ultranarrow {F}araday rotation filter at the {R}b {D$_1$} line},}\ }\href {\doibase 10.1364/OL.37.000524} {\bibfield  {journal} {\bibinfo  {journal} {Opt. Lett.}\ }\textbf {\bibinfo {volume} {37}},\ \bibinfo {pages} {524--526} (\bibinfo {year} {2012})}\BibitemShut {NoStop}%
\bibitem [{\citenamefont {Zieli\'{n}ska}\ \emph {et~al.}(2014)\citenamefont {Zieli\'{n}ska}, \citenamefont {Beduini}, \citenamefont {Lucivero},\ and\ \citenamefont {Mitchell}}]{ZielinskaOE2014}%
  \BibitemOpen
  \bibfield  {author} {\bibinfo {author} {\bibfnamefont {Joanna~A.}\ \bibnamefont {Zieli\'{n}ska}}, \bibinfo {author} {\bibfnamefont {Federica~A.}\ \bibnamefont {Beduini}}, \bibinfo {author} {\bibfnamefont {Vito~Giovanni}\ \bibnamefont {Lucivero}}, \ and\ \bibinfo {author} {\bibfnamefont {Morgan~W.}\ \bibnamefont {Mitchell}},\ }\bibfield  {title} {\enquote {\bibinfo {title} {Atomic filtering for hybrid continuous-variable/discrete-variable quantum optics},}\ }\href {\doibase 10.1364/OE.22.025307} {\bibfield  {journal} {\bibinfo  {journal} {Opt. Express}\ }\textbf {\bibinfo {volume} {22}},\ \bibinfo {pages} {25307--25317} (\bibinfo {year} {2014})}\BibitemShut {NoStop}%
\bibitem [{\citenamefont {Xia}\ \emph {et~al.}(2023)\citenamefont {Xia}, \citenamefont {Cheng}, \citenamefont {Wang}, \citenamefont {Liu}, \citenamefont {Yang}, \citenamefont {Du}, \citenamefont {Jiao}, \citenamefont {Wang}, \citenamefont {Zheng}, \citenamefont {Li}, \citenamefont {Li},\ and\ \citenamefont {Yang}}]{XiaRS2023}%
  \BibitemOpen
  \bibfield  {author} {\bibinfo {author} {\bibfnamefont {Yuan}\ \bibnamefont {Xia}}, \bibinfo {author} {\bibfnamefont {Xuewu}\ \bibnamefont {Cheng}}, \bibinfo {author} {\bibfnamefont {Zelong}\ \bibnamefont {Wang}}, \bibinfo {author} {\bibfnamefont {Linmei}\ \bibnamefont {Liu}}, \bibinfo {author} {\bibfnamefont {Yong}\ \bibnamefont {Yang}}, \bibinfo {author} {\bibfnamefont {Lifang}\ \bibnamefont {Du}}, \bibinfo {author} {\bibfnamefont {Jing}\ \bibnamefont {Jiao}}, \bibinfo {author} {\bibfnamefont {Jihong}\ \bibnamefont {Wang}}, \bibinfo {author} {\bibfnamefont {Haoran}\ \bibnamefont {Zheng}}, \bibinfo {author} {\bibfnamefont {Yajuan}\ \bibnamefont {Li}}, \bibinfo {author} {\bibfnamefont {Faquan}\ \bibnamefont {Li}}, \ and\ \bibinfo {author} {\bibfnamefont {Guotao}\ \bibnamefont {Yang}},\ }\bibfield  {title} {\enquote {\bibinfo {title} {Design of a data acquisition, correction and retrieval of {Na} {Doppler} lidar for diurnal measurement of temperature and wind in the mesosphere and lower thermosphere
  region},}\ }\href {\doibase 10.3390/rs15215140} {\bibfield  {journal} {\bibinfo  {journal} {Remote Sensing}\ }\textbf {\bibinfo {volume} {15}} (\bibinfo {year} {2023}),\ 10.3390/rs15215140}\BibitemShut {NoStop}%
\bibitem [{Note1()}]{Note1}%
  \BibitemOpen
  \bibinfo {note} {EIT systems \cite {Zou2017} can have narrower transmission windows within a narrow blocking band. \SI {}{\giga \hertz }-bandwidth atomic vapor filters are used in rejection of daylight background \cite {KieferSR2014, XiaRS2023} and can outperform classical filters in some contexts \cite { YinIEEEPJ}. A relevant measure is the equivalent noise bandwidth \cite {ZielinskaOL2012}.}\BibitemShut {Stop}%
\bibitem [{\citenamefont {Bowden}\ \emph {et~al.}(2019)\citenamefont {Bowden}, \citenamefont {Hobson}, \citenamefont {Hill}, \citenamefont {Vianello}, \citenamefont {Schioppo}, \citenamefont {Silva}, \citenamefont {Margolis}, \citenamefont {Baird},\ and\ \citenamefont {Gill}}]{BowdenSR2019}%
  \BibitemOpen
  \bibfield  {author} {\bibinfo {author} {\bibfnamefont {William}\ \bibnamefont {Bowden}}, \bibinfo {author} {\bibfnamefont {Richard}\ \bibnamefont {Hobson}}, \bibinfo {author} {\bibfnamefont {Ian~R.}\ \bibnamefont {Hill}}, \bibinfo {author} {\bibfnamefont {Alvise}\ \bibnamefont {Vianello}}, \bibinfo {author} {\bibfnamefont {Marco}\ \bibnamefont {Schioppo}}, \bibinfo {author} {\bibfnamefont {Alissa}\ \bibnamefont {Silva}}, \bibinfo {author} {\bibfnamefont {Helen~S.}\ \bibnamefont {Margolis}}, \bibinfo {author} {\bibfnamefont {Patrick E.~G.}\ \bibnamefont {Baird}}, \ and\ \bibinfo {author} {\bibfnamefont {Patrick}\ \bibnamefont {Gill}},\ }\bibfield  {title} {\enquote {\bibinfo {title} {A pyramid {MOT} with integrated optical cavities as a cold atom platform for an optical lattice clock},}\ }\href {\doibase 10.1038/s41598-019-48168-3} {\bibfield  {journal} {\bibinfo  {journal} {Scientific Reports}\ }\textbf {\bibinfo {volume} {9}},\ \bibinfo {pages} {11704} (\bibinfo {year} {2019})}\BibitemShut {NoStop}%
\bibitem [{\citenamefont {Wang}\ \emph {et~al.}(2023)\citenamefont {Wang}, \citenamefont {Jiao}, \citenamefont {Wang}, \citenamefont {Liu}, \citenamefont {Xie}, \citenamefont {Zheng}, \citenamefont {Zhang},\ and\ \citenamefont {Pan}}]{WangNPJQI2023}%
  \BibitemOpen
  \bibfield  {author} {\bibinfo {author} {\bibfnamefont {Xina}\ \bibnamefont {Wang}}, \bibinfo {author} {\bibfnamefont {Xufeng}\ \bibnamefont {Jiao}}, \bibinfo {author} {\bibfnamefont {Bin}\ \bibnamefont {Wang}}, \bibinfo {author} {\bibfnamefont {Yang}\ \bibnamefont {Liu}}, \bibinfo {author} {\bibfnamefont {Xiu-Ping}\ \bibnamefont {Xie}}, \bibinfo {author} {\bibfnamefont {Ming-Yang}\ \bibnamefont {Zheng}}, \bibinfo {author} {\bibfnamefont {Qiang}\ \bibnamefont {Zhang}}, \ and\ \bibinfo {author} {\bibfnamefont {Jian-Wei}\ \bibnamefont {Pan}},\ }\bibfield  {title} {\enquote {\bibinfo {title} {Quantum frequency conversion and single-photon detection with lithium niobate nanophotonic chips},}\ }\href {\doibase 10.1038/s41534-023-00704-w} {\bibfield  {journal} {\bibinfo  {journal} {npj Quantum Information}\ }\textbf {\bibinfo {volume} {9}},\ \bibinfo {pages} {38} (\bibinfo {year} {2023})}\BibitemShut {NoStop}%
\bibitem [{\citenamefont {Tey}\ \emph {et~al.}(2009)\citenamefont {Tey}, \citenamefont {Maslennikov}, \citenamefont {Liew}, \citenamefont {Aljunid}, \citenamefont {Huber}, \citenamefont {Chng}, \citenamefont {Chen}, \citenamefont {Scarani},\ and\ \citenamefont {Kurtsiefer}}]{TeyNJP2009}%
  \BibitemOpen
  \bibfield  {author} {\bibinfo {author} {\bibfnamefont {Meng~Khoon}\ \bibnamefont {Tey}}, \bibinfo {author} {\bibfnamefont {Gleb}\ \bibnamefont {Maslennikov}}, \bibinfo {author} {\bibfnamefont {Timothy C~H}\ \bibnamefont {Liew}}, \bibinfo {author} {\bibfnamefont {Syed~Abdullah}\ \bibnamefont {Aljunid}}, \bibinfo {author} {\bibfnamefont {Florian}\ \bibnamefont {Huber}}, \bibinfo {author} {\bibfnamefont {Brenda}\ \bibnamefont {Chng}}, \bibinfo {author} {\bibfnamefont {Zilong}\ \bibnamefont {Chen}}, \bibinfo {author} {\bibfnamefont {Valerio}\ \bibnamefont {Scarani}}, \ and\ \bibinfo {author} {\bibfnamefont {Christian}\ \bibnamefont {Kurtsiefer}},\ }\bibfield  {title} {\enquote {\bibinfo {title} {Interfacing light and single atoms with a lens},}\ }\href {http://stacks.iop.org/1367-2630/11/i=4/a=043011} {\bibfield  {journal} {\bibinfo  {journal} {New Journal of Physics}\ }\textbf {\bibinfo {volume} {11}},\ \bibinfo {pages} {043011} (\bibinfo {year} {2009})}\BibitemShut {NoStop}%
\bibitem [{\citenamefont {Sondermann}\ \emph {et~al.}(2020)\citenamefont {Sondermann}, \citenamefont {Fischer},\ and\ \citenamefont {Leuchs}}]{SondermannAQT2020}%
  \BibitemOpen
  \bibfield  {author} {\bibinfo {author} {\bibfnamefont {Markus}\ \bibnamefont {Sondermann}}, \bibinfo {author} {\bibfnamefont {Martin}\ \bibnamefont {Fischer}}, \ and\ \bibinfo {author} {\bibfnamefont {Gerd}\ \bibnamefont {Leuchs}},\ }\bibfield  {title} {\enquote {\bibinfo {title} {Prospects of trapping atoms with an optical dipole trap in a deep parabolic mirror for light--matter-interaction experiments},}\ }\href {\doibase 10.1002/qute.202000022} {\bibfield  {journal} {\bibinfo  {journal} {Advanced Quantum Technologies}\ }\textbf {\bibinfo {volume} {3}},\ \bibinfo {pages} {2000022} (\bibinfo {year} {2020})}\BibitemShut {NoStop}%
\bibitem [{\citenamefont {Nguyen}\ \emph {et~al.}(2017)\citenamefont {Nguyen}, \citenamefont {Utama}, \citenamefont {Lewty}, \citenamefont {Durak}, \citenamefont {Maslennikov}, \citenamefont {Straupe}, \citenamefont {Steiner},\ and\ \citenamefont {Kurtsiefer}}]{NguyenPRA2017}%
  \BibitemOpen
  \bibfield  {author} {\bibinfo {author} {\bibfnamefont {Chi~Huan}\ \bibnamefont {Nguyen}}, \bibinfo {author} {\bibfnamefont {Adrian~Nugraha}\ \bibnamefont {Utama}}, \bibinfo {author} {\bibfnamefont {Nick}\ \bibnamefont {Lewty}}, \bibinfo {author} {\bibfnamefont {Kadir}\ \bibnamefont {Durak}}, \bibinfo {author} {\bibfnamefont {Gleb}\ \bibnamefont {Maslennikov}}, \bibinfo {author} {\bibfnamefont {Stanislav}\ \bibnamefont {Straupe}}, \bibinfo {author} {\bibfnamefont {Matthias}\ \bibnamefont {Steiner}}, \ and\ \bibinfo {author} {\bibfnamefont {Christian}\ \bibnamefont {Kurtsiefer}},\ }\bibfield  {title} {\enquote {\bibinfo {title} {Single atoms coupled to a near-concentric cavity},}\ }\href {\doibase 10.1103/PhysRevA.96.031802} {\bibfield  {journal} {\bibinfo  {journal} {Phys. Rev. A}\ }\textbf {\bibinfo {volume} {96}},\ \bibinfo {pages} {0318027} (\bibinfo {year} {2017})}\BibitemShut {NoStop}%
\bibitem [{\citenamefont {Nagourney}\ \emph {et~al.}(1986)\citenamefont {Nagourney}, \citenamefont {Sandberg},\ and\ \citenamefont {Dehmelt}}]{NagourneyPRL1986}%
  \BibitemOpen
  \bibfield  {author} {\bibinfo {author} {\bibfnamefont {Warren}\ \bibnamefont {Nagourney}}, \bibinfo {author} {\bibfnamefont {Jon}\ \bibnamefont {Sandberg}}, \ and\ \bibinfo {author} {\bibfnamefont {Hans}\ \bibnamefont {Dehmelt}},\ }\bibfield  {title} {\enquote {\bibinfo {title} {Shelved optical electron amplifier: Observation of quantum jumps},}\ }\href {\doibase 10.1103/PhysRevLett.56.2797} {\bibfield  {journal} {\bibinfo  {journal} {Phys. Rev. Lett.}\ }\textbf {\bibinfo {volume} {56}},\ \bibinfo {pages} {2797--2799} (\bibinfo {year} {1986})}\BibitemShut {NoStop}%
\bibitem [{\citenamefont {Bianchet}\ \emph {et~al.}(2022)\citenamefont {Bianchet}, \citenamefont {Alves}, \citenamefont {Zarraoa}, \citenamefont {Lamich}, \citenamefont {Prakash},\ and\ \citenamefont {Mitchell}}]{Bianchet2022}%
  \BibitemOpen
  \bibfield  {author} {\bibinfo {author} {\bibfnamefont {Lorena~C.}\ \bibnamefont {Bianchet}}, \bibinfo {author} {\bibfnamefont {Natalia}\ \bibnamefont {Alves}}, \bibinfo {author} {\bibfnamefont {Laura}\ \bibnamefont {Zarraoa}}, \bibinfo {author} {\bibfnamefont {Tomas}\ \bibnamefont {Lamich}}, \bibinfo {author} {\bibfnamefont {Vindhiya}\ \bibnamefont {Prakash}}, \ and\ \bibinfo {author} {\bibfnamefont {Morgan~W.}\ \bibnamefont {Mitchell}},\ }\bibfield  {title} {\enquote {\bibinfo {title} {Quantum jump spectroscopy of a single neutral atom for precise subwavelength intensity measurements},}\ }\href {\doibase 10.1103/PhysRevResearch.4.L042026} {\bibfield  {journal} {\bibinfo  {journal} {Phys. Rev. Res.}\ }\textbf {\bibinfo {volume} {4}},\ \bibinfo {pages} {L042026} (\bibinfo {year} {2022})}\BibitemShut {NoStop}%
\bibitem [{\citenamefont {Aljunid}\ \emph {et~al.}(2013)\citenamefont {Aljunid}, \citenamefont {Maslennikov}, \citenamefont {Wang}, \citenamefont {Dao}, \citenamefont {Scarani},\ and\ \citenamefont {Kurtsiefer}}]{AljunidPRL2013}%
  \BibitemOpen
  \bibfield  {author} {\bibinfo {author} {\bibfnamefont {Syed~Abdullah}\ \bibnamefont {Aljunid}}, \bibinfo {author} {\bibfnamefont {Gleb}\ \bibnamefont {Maslennikov}}, \bibinfo {author} {\bibfnamefont {Yimin}\ \bibnamefont {Wang}}, \bibinfo {author} {\bibfnamefont {Hoang~Lan}\ \bibnamefont {Dao}}, \bibinfo {author} {\bibfnamefont {Valerio}\ \bibnamefont {Scarani}}, \ and\ \bibinfo {author} {\bibfnamefont {Christian}\ \bibnamefont {Kurtsiefer}},\ }\bibfield  {title} {\enquote {\bibinfo {title} {Excitation of a single atom with exponentially rising light pulses},}\ }\href {\doibase 10.1103/PhysRevLett.111.103001} {\bibfield  {journal} {\bibinfo  {journal} {Phys. Rev. Lett.}\ }\textbf {\bibinfo {volume} {111}},\ \bibinfo {pages} {103001} (\bibinfo {year} {2013})}\BibitemShut {NoStop}%
\bibitem [{\citenamefont {Bruno}\ \emph {et~al.}(2019)\citenamefont {Bruno}, \citenamefont {Bianchet}, \citenamefont {Prakash}, \citenamefont {Li}, \citenamefont {Alves},\ and\ \citenamefont {Mitchell}}]{BrunoOE2019}%
  \BibitemOpen
  \bibfield  {author} {\bibinfo {author} {\bibfnamefont {Natalia}\ \bibnamefont {Bruno}}, \bibinfo {author} {\bibfnamefont {Lorena~C.}\ \bibnamefont {Bianchet}}, \bibinfo {author} {\bibfnamefont {Vindhiya}\ \bibnamefont {Prakash}}, \bibinfo {author} {\bibfnamefont {Nan}\ \bibnamefont {Li}}, \bibinfo {author} {\bibfnamefont {Nat\'{a}lia}\ \bibnamefont {Alves}}, \ and\ \bibinfo {author} {\bibfnamefont {Morgan~W.}\ \bibnamefont {Mitchell}},\ }\bibfield  {title} {\enquote {\bibinfo {title} {Maltese cross coupling to individual cold atoms in free space},}\ }\href {\doibase 10.1364/OE.27.031042} {\bibfield  {journal} {\bibinfo  {journal} {Opt. Express}\ }\textbf {\bibinfo {volume} {27}},\ \bibinfo {pages} {31042--31052} (\bibinfo {year} {2019})}\BibitemShut {NoStop}%
\bibitem [{\citenamefont {Bianchet}\ \emph {et~al.}(2021)\citenamefont {Bianchet}, \citenamefont {Alves}, \citenamefont {Zarraoa}, \citenamefont {Bruno},\ and\ \citenamefont {Mitchell}}]{BianchetORE2021}%
  \BibitemOpen
  \bibfield  {author} {\bibinfo {author} {\bibfnamefont {L.C.}\ \bibnamefont {Bianchet}}, \bibinfo {author} {\bibfnamefont {N.}~\bibnamefont {Alves}}, \bibinfo {author} {\bibfnamefont {L.}~\bibnamefont {Zarraoa}}, \bibinfo {author} {\bibfnamefont {N.}~\bibnamefont {Bruno}}, \ and\ \bibinfo {author} {\bibfnamefont {M.W.}\ \bibnamefont {Mitchell}},\ }\bibfield  {title} {\enquote {\bibinfo {title} {Manipulating and measuring single atoms in the {Maltese} cross geometry},}\ }\href {\doibase 10.12688/openreseurope.13972.1} {\bibfield  {journal} {\bibinfo  {journal} {Open Research Europe}\ }\textbf {\bibinfo {volume} {1}} (\bibinfo {year} {2021}),\ 10.12688/openreseurope.13972.1}\BibitemShut {NoStop}%
\bibitem [{\citenamefont {Coop}\ \emph {et~al.}(2017)\citenamefont {Coop}, \citenamefont {Palacios}, \citenamefont {Gomez}, \citenamefont {de~Escobar}, \citenamefont {Vanderbruggen},\ and\ \citenamefont {Mitchell}}]{SimonCoopLightshift2017}%
  \BibitemOpen
  \bibfield  {author} {\bibinfo {author} {\bibfnamefont {Simon}\ \bibnamefont {Coop}}, \bibinfo {author} {\bibfnamefont {Silvana}\ \bibnamefont {Palacios}}, \bibinfo {author} {\bibfnamefont {Pau}\ \bibnamefont {Gomez}}, \bibinfo {author} {\bibfnamefont {Y.~Natali~Martinez}\ \bibnamefont {de~Escobar}}, \bibinfo {author} {\bibfnamefont {Thomas}\ \bibnamefont {Vanderbruggen}}, \ and\ \bibinfo {author} {\bibfnamefont {Morgan~W.}\ \bibnamefont {Mitchell}},\ }\bibfield  {title} {\enquote {\bibinfo {title} {Floquet theory for atomic light-shift engineering with near-resonant polychromatic fields},}\ }\href {\doibase 10.1364/OE.25.032550} {\bibfield  {journal} {\bibinfo  {journal} {Opt. Express}\ }\textbf {\bibinfo {volume} {25}},\ \bibinfo {pages} {32550--32559} (\bibinfo {year} {2017})}\BibitemShut {NoStop}%
\bibitem [{\citenamefont {Peterson}(2017)}]{PetersonBook2017}%
  \BibitemOpen
  \bibfield  {author} {\bibinfo {author} {\bibfnamefont {M.}~\bibnamefont {Peterson}},\ }\href {https://books.google.es/books?id=CnQ3DgAAQBAJ} {\emph {\bibinfo {title} {An Introduction to Decision Theory}}}\ (\bibinfo  {publisher} {Cambridge University Press},\ \bibinfo {year} {2017})\BibitemShut {NoStop}%
\bibitem [{\citenamefont {Fuhrmanek}\ \emph {et~al.}(2011)\citenamefont {Fuhrmanek}, \citenamefont {Bourgain}, \citenamefont {Sortais},\ and\ \citenamefont {Browaeys}}]{Fuhrmanek2011}%
  \BibitemOpen
  \bibfield  {author} {\bibinfo {author} {\bibfnamefont {A.}~\bibnamefont {Fuhrmanek}}, \bibinfo {author} {\bibfnamefont {R.}~\bibnamefont {Bourgain}}, \bibinfo {author} {\bibfnamefont {Y.~R.~P.}\ \bibnamefont {Sortais}}, \ and\ \bibinfo {author} {\bibfnamefont {A.}~\bibnamefont {Browaeys}},\ }\bibfield  {title} {\enquote {\bibinfo {title} {Free-space lossless state detection of a single trapped atom},}\ }\href {\doibase 10.1103/PhysRevLett.106.133003} {\bibfield  {journal} {\bibinfo  {journal} {Phys. Rev. Lett.}\ }\textbf {\bibinfo {volume} {106}},\ \bibinfo {pages} {133003} (\bibinfo {year} {2011})}\BibitemShut {NoStop}%
\bibitem [{\citenamefont {Hemmerling}\ \emph {et~al.}(2012)\citenamefont {Hemmerling}, \citenamefont {Gebert}, \citenamefont {Wan},\ and\ \citenamefont {Schmidt}}]{Hemmerling2012}%
  \BibitemOpen
  \bibfield  {author} {\bibinfo {author} {\bibfnamefont {B}~\bibnamefont {Hemmerling}}, \bibinfo {author} {\bibfnamefont {F}~\bibnamefont {Gebert}}, \bibinfo {author} {\bibfnamefont {Y}~\bibnamefont {Wan}}, \ and\ \bibinfo {author} {\bibfnamefont {P~O}\ \bibnamefont {Schmidt}},\ }\bibfield  {title} {\enquote {\bibinfo {title} {A novel, robust quantum detection scheme},}\ }\href {\doibase 10.1088/1367-2630/14/2/023043} {\bibfield  {journal} {\bibinfo  {journal} {New Journal of Physics}\ }\textbf {\bibinfo {volume} {14}},\ \bibinfo {pages} {023043} (\bibinfo {year} {2012})}\BibitemShut {NoStop}%
\bibitem [{\citenamefont {Volz}\ \emph {et~al.}(2006)\citenamefont {Volz}, \citenamefont {Weber}, \citenamefont {Schlenk}, \citenamefont {Rosenfeld}, \citenamefont {Vrana}, \citenamefont {Saucke}, \citenamefont {Kurtsiefer},\ and\ \citenamefont {Weinfurter}}]{VolzWeberSchlenkEtAl2006}%
  \BibitemOpen
  \bibfield  {author} {\bibinfo {author} {\bibfnamefont {J{\"u}rgen}\ \bibnamefont {Volz}}, \bibinfo {author} {\bibfnamefont {Markus}\ \bibnamefont {Weber}}, \bibinfo {author} {\bibfnamefont {Daniel}\ \bibnamefont {Schlenk}}, \bibinfo {author} {\bibfnamefont {Wenjamin}\ \bibnamefont {Rosenfeld}}, \bibinfo {author} {\bibfnamefont {Johannes}\ \bibnamefont {Vrana}}, \bibinfo {author} {\bibfnamefont {Karen}\ \bibnamefont {Saucke}}, \bibinfo {author} {\bibfnamefont {Christian}\ \bibnamefont {Kurtsiefer}}, \ and\ \bibinfo {author} {\bibfnamefont {Harald}\ \bibnamefont {Weinfurter}},\ }\bibfield  {title} {\enquote {\bibinfo {title} {Observation of entanglement of a single photon with a trapped atom},}\ }\href {\doibase 10.1103/PhysRevLett.96.030404} {\bibfield  {journal} {\bibinfo  {journal} {Phys. Rev. Lett.}\ }\textbf {\bibinfo {volume} {96}},\ \bibinfo {pages} {030404} (\bibinfo {year} {2006})}\BibitemShut {NoStop}%
\bibitem [{\citenamefont {Hamamatsu}(2022)}]{SAPDs}%
  \BibitemOpen
  \bibfield  {author} {\bibinfo {author} {\bibnamefont {Hamamatsu}},\ }\bibfield  {title} {\enquote {\bibinfo {title} {Photon counting modules {C11202} series: {C11202-050}},}\ }\href {https://www.hamamatsu.com/content/dam/hamamatsu-photonics/sites/documents/99_SALES_LIBRARY/ssd/c11202series_kacc1207e.pdf} {\bibfield  {journal} {\bibinfo  {journal} {Datasheet Cat. No. KACC1207E07 Sep. 2022 DN}\ } (\bibinfo {year} {2022})}\BibitemShut {NoStop}%
\bibitem [{\citenamefont {Shaw}(2018)}]{SiNWs_Psyche}%
  \BibitemOpen
  \bibfield  {author} {\bibinfo {author} {\bibfnamefont {Matt}\ \bibnamefont {Shaw}},\ }\bibfield  {title} {\enquote {\bibinfo {title} {Superconducting nanowire single photon detectors for deep space optical communication},}\ }\href {https://indico.physics.lbl.gov/event/815/attachments/1750/2119/APH_110_2018.pdf} {\bibfield  {journal} {\bibinfo  {journal} {683rd WE-Heraeus-Seminar}\ } (\bibinfo {year} {2018})}\BibitemShut {NoStop}%
\bibitem [{\citenamefont {Zhang}\ \emph {et~al.}(2022)\citenamefont {Zhang}, \citenamefont {Tiwari}, \citenamefont {Ganesh}, \citenamefont {Ramchurn}, \citenamefont {Bongs},\ and\ \citenamefont {Singh}}]{ZhangEFTFIFCS2022}%
  \BibitemOpen
  \bibfield  {author} {\bibinfo {author} {\bibfnamefont {Shengnan}\ \bibnamefont {Zhang}}, \bibinfo {author} {\bibfnamefont {Balsant}\ \bibnamefont {Tiwari}}, \bibinfo {author} {\bibfnamefont {Sandhya}\ \bibnamefont {Ganesh}}, \bibinfo {author} {\bibfnamefont {Preetam}\ \bibnamefont {Ramchurn}}, \bibinfo {author} {\bibfnamefont {Kai}\ \bibnamefont {Bongs}}, \ and\ \bibinfo {author} {\bibfnamefont {Yeshpal}\ \bibnamefont {Singh}},\ }\bibfield  {title} {\enquote {\bibinfo {title} {Blue-detuned optical lattice for sr long-range interactions},}\ }in\ \href {\doibase 10.1109/EFTF/IFCS54560.2022.9850544} {\emph {\bibinfo {booktitle} {2022 Joint Conference of the European Frequency and Time Forum and IEEE International Frequency Control Symposium (EFTF/IFCS)}}}\ (\bibinfo {year} {2022})\ pp.\ \bibinfo {pages} {1--3}\BibitemShut {NoStop}%
\bibitem [{\citenamefont {Piro}\ \emph {et~al.}(2011)\citenamefont {Piro}, \citenamefont {Rohde}, \citenamefont {Schuck}, \citenamefont {Almendros}, \citenamefont {Huwer}, \citenamefont {Ghosh}, \citenamefont {Haase}, \citenamefont {Hennrich}, \citenamefont {Dubin},\ and\ \citenamefont {Eschner}}]{PiroNP2011}%
  \BibitemOpen
  \bibfield  {author} {\bibinfo {author} {\bibfnamefont {N.}~\bibnamefont {Piro}}, \bibinfo {author} {\bibfnamefont {F.}~\bibnamefont {Rohde}}, \bibinfo {author} {\bibfnamefont {C.}~\bibnamefont {Schuck}}, \bibinfo {author} {\bibfnamefont {M.}~\bibnamefont {Almendros}}, \bibinfo {author} {\bibfnamefont {J.}~\bibnamefont {Huwer}}, \bibinfo {author} {\bibfnamefont {J.}~\bibnamefont {Ghosh}}, \bibinfo {author} {\bibfnamefont {A.}~\bibnamefont {Haase}}, \bibinfo {author} {\bibfnamefont {M.}~\bibnamefont {Hennrich}}, \bibinfo {author} {\bibfnamefont {F.}~\bibnamefont {Dubin}}, \ and\ \bibinfo {author} {\bibfnamefont {J.}~\bibnamefont {Eschner}},\ }\bibfield  {title} {\enquote {\bibinfo {title} {Heralded single-photon absorption by a single atom},}\ }\href {http://dx.doi.org/10.1038/nphys1805} {\bibfield  {journal} {\bibinfo  {journal} {Nat Phys}\ }\textbf {\bibinfo {volume} {7}},\ \bibinfo {pages} {17--20} (\bibinfo {year} {2011})}\BibitemShut {NoStop}%
\bibitem [{\citenamefont {Specht}\ \emph {et~al.}(2011)\citenamefont {Specht}, \citenamefont {Nolleke}, \citenamefont {Reiserer}, \citenamefont {Uphoff}, \citenamefont {Figueroa}, \citenamefont {Ritter},\ and\ \citenamefont {Rempe}}]{SpechtN2011}%
  \BibitemOpen
  \bibfield  {author} {\bibinfo {author} {\bibfnamefont {Holger~P.}\ \bibnamefont {Specht}}, \bibinfo {author} {\bibfnamefont {Christian}\ \bibnamefont {Nolleke}}, \bibinfo {author} {\bibfnamefont {Andreas}\ \bibnamefont {Reiserer}}, \bibinfo {author} {\bibfnamefont {Manuel}\ \bibnamefont {Uphoff}}, \bibinfo {author} {\bibfnamefont {Eden}\ \bibnamefont {Figueroa}}, \bibinfo {author} {\bibfnamefont {Stephan}\ \bibnamefont {Ritter}}, \ and\ \bibinfo {author} {\bibfnamefont {Gerhard}\ \bibnamefont {Rempe}},\ }\bibfield  {title} {\enquote {\bibinfo {title} {A single-atom quantum memory},}\ }\href {http://dx.doi.org/10.1038/nature09997} {\bibfield  {journal} {\bibinfo  {journal} {Nature}\ }\textbf {\bibinfo {volume} {473}},\ \bibinfo {pages} {190--193} (\bibinfo {year} {2011})}\BibitemShut {NoStop}%
\bibitem [{\citenamefont {Kuhn}\ \emph {et~al.}(2002)\citenamefont {Kuhn}, \citenamefont {Hennrich},\ and\ \citenamefont {Rempe}}]{KuhnPRL2002}%
  \BibitemOpen
  \bibfield  {author} {\bibinfo {author} {\bibfnamefont {Axel}\ \bibnamefont {Kuhn}}, \bibinfo {author} {\bibfnamefont {Markus}\ \bibnamefont {Hennrich}}, \ and\ \bibinfo {author} {\bibfnamefont {Gerhard}\ \bibnamefont {Rempe}},\ }\bibfield  {title} {\enquote {\bibinfo {title} {Deterministic single-photon source for distributed quantum networking},}\ }\href {\doibase 10.1103/PhysRevLett.89.067901} {\bibfield  {journal} {\bibinfo  {journal} {Physical Review Letters}\ }\textbf {\bibinfo {volume} {89}},\ \bibinfo {pages} {067901} (\bibinfo {year} {2002})}\BibitemShut {NoStop}%
\bibitem [{\citenamefont {Zou}\ \emph {et~al.}(2017)\citenamefont {Zou}, \citenamefont {Jiang}, \citenamefont {Mei}, \citenamefont {Guo},\ and\ \citenamefont {Du}}]{Zou2017}%
  \BibitemOpen
  \bibfield  {author} {\bibinfo {author} {\bibfnamefont {Yueyang}\ \bibnamefont {Zou}}, \bibinfo {author} {\bibfnamefont {Yue}\ \bibnamefont {Jiang}}, \bibinfo {author} {\bibfnamefont {Yefeng}\ \bibnamefont {Mei}}, \bibinfo {author} {\bibfnamefont {Xianxin}\ \bibnamefont {Guo}}, \ and\ \bibinfo {author} {\bibfnamefont {Shengwang}\ \bibnamefont {Du}},\ }\bibfield  {title} {\enquote {\bibinfo {title} {Quantum heat engine using electromagnetically induced transparency},}\ }\href {\doibase 10.1103/PhysRevLett.119.050602} {\bibfield  {journal} {\bibinfo  {journal} {Phys. Rev. Lett.}\ }\textbf {\bibinfo {volume} {119}},\ \bibinfo {pages} {050602} (\bibinfo {year} {2017})}\BibitemShut {NoStop}%
\bibitem [{\citenamefont {Kiefer}\ \emph {et~al.}(2014)\citenamefont {Kiefer}, \citenamefont {L\"{o}w}, \citenamefont {Wrachtrup},\ and\ \citenamefont {Gerhardt}}]{KieferSR2014}%
  \BibitemOpen
  \bibfield  {author} {\bibinfo {author} {\bibfnamefont {Wilhelm}\ \bibnamefont {Kiefer}}, \bibinfo {author} {\bibfnamefont {Robert}\ \bibnamefont {L\"{o}w}}, \bibinfo {author} {\bibfnamefont {J\"{o}rg}\ \bibnamefont {Wrachtrup}}, \ and\ \bibinfo {author} {\bibfnamefont {Ilja}\ \bibnamefont {Gerhardt}},\ }\bibfield  {title} {\enquote {\bibinfo {title} {Na-{Faraday} rotation filtering: {T}he optimal point},}\ }\href {\doibase 10.1038/srep06552} {\bibfield  {journal} {\bibinfo  {journal} {Scientific Reports}\ }\textbf {\bibinfo {volume} {4}},\ \bibinfo {pages} {6552} (\bibinfo {year} {2014})}\BibitemShut {NoStop}%
\bibitem [{\citenamefont {Yin}\ \emph {et~al.}(2024)\citenamefont {Yin}, \citenamefont {Tang}, \citenamefont {Geng}, \citenamefont {Zhan}, \citenamefont {Chen}, \citenamefont {Qian}, \citenamefont {Wu},\ and\ \citenamefont {Luo}}]{YinIEEEPJ}%
  \BibitemOpen
  \bibfield  {author} {\bibinfo {author} {\bibfnamefont {Longfei}\ \bibnamefont {Yin}}, \bibinfo {author} {\bibfnamefont {Wei}\ \bibnamefont {Tang}}, \bibinfo {author} {\bibfnamefont {Ziwei}\ \bibnamefont {Geng}}, \bibinfo {author} {\bibfnamefont {Haodi}\ \bibnamefont {Zhan}}, \bibinfo {author} {\bibfnamefont {Lei}\ \bibnamefont {Chen}}, \bibinfo {author} {\bibfnamefont {Dasheng}\ \bibnamefont {Qian}}, \bibinfo {author} {\bibfnamefont {Guohua}\ \bibnamefont {Wu}}, \ and\ \bibinfo {author} {\bibfnamefont {Bin}\ \bibnamefont {Luo}},\ }\bibfield  {title} {\enquote {\bibinfo {title} {Sunlight noise mitigation in fmcw lidar using fadof},}\ }\href {\doibase 10.1109/JPHOT.2024.3388326} {\bibfield  {journal} {\bibinfo  {journal} {IEEE Photonics Journal}\ }\textbf {\bibinfo {volume} {16}},\ \bibinfo {pages} {1--8} (\bibinfo {year} {2024})}\BibitemShut {NoStop}%
\end{thebibliography}

\end{document}